\documentclass[manuscript]{emulateapj}

\usepackage{multirow}
\usepackage{graphicx}
\usepackage{amsmath, amsthm, amssymb}
\usepackage{xcolor}

\begin{document}

\title{Kiloparsec-Scale Simulations of Star Formation in Disk Galaxies.\\ I.
The unmagnetized and zero-feedback limit.}
\author{Sven Van Loo\altaffilmark{1}}
\affil{Harvard-Smithsonian Center for Astrophysics, 60 Garden Street, Cambridge, MA 02138, USA}
\altaffiltext{1}{Department of Astronomy, University of Florida, Gainesville, FL 32611, USA}
\email{svanloo@cfa.harvard.edu}
\author{Michael J. Butler,  Jonathan C. Tan}
\affil{Department of Astronomy, University of Florida, Gainesville, FL 32611, USA}

\begin{abstract}
We present hydrodynamic simulations of the evolution of self-gravitating dense gas 
on scales of 1~kiloparsec down to $\lesssim$parsec in a galactic disk, 
designed to study dense clump formation from giant molecular clouds (GMCs). These 
structures are expected to be the precursors to star clusters and this process may 
be the rate limiting step controling star formation rates in galactic systems as
described by the Kennicutt-Schmidt relation. We follow the thermal evolution of the 
gas down to $\sim 5$~K using extinction-dependent heating and cooling 
functions. We do not yet include magnetic fields or localized stellar feedback, so 
the evolution of the GMCs and clumps is determined solely by 
self-gravity balanced by thermal and turbulent pressure support and the large scale 
galactic shear.  While cloud structures and densities change significantly 
during the simulation, GMC virial parameters remain mostly above unity for time
scales exceeding the free-fall time of GMCs indicating that energy from galactic shear 
and large-scale cloud motions continuously cascades down to and within the GMCs. We 
implement star formation at a slow, inefficient rate of 2\% per local free-fall time, 
but even this yields global star formation rates that are about two orders of magnitude 
larger than the observed Kennicutt-Schmidt relation due to over-production 
of dense gas clumps. We expect a combination of magnetic support and localized stellar 
feedback is required to inhibit dense clump formation to $\sim$1\% of the rate that 
results from the nonmagnetic, zero-feedback limit.
\end{abstract}


\maketitle
\section{Introduction}
Star formation in galaxies involves a vast range of length and time scales, from the 
tens of kiloparsec diameters and $\sim 10^8$~yr orbits of galactic disks to the 
$\sim 0.1$~pc sizes and $\sim 10^5$~yr dynamical times of individual prestellar cores 
(PSCs) \citep[see][ for reviews]{McKeeOstriker2007,ScaloElmegreen2004,
ElmegreenScalo2004,MacLowKlessen2004,Ballesterosetal2007}. Self-gravity in the gas is 
effectively countered by various forms of pressure support (including thermal, 
magnetic and turbulent), large-scale coherent motions (including galactic 
shear and large-scale turbulent flows) that drive turbulent motions, and localized 
feedback from newborn stars in order to make the overall star formation rate relatively 
slow and inefficient at just a few percent conversion of gas to stars per local 
dynamical timescale across a wide range of densities \citep[][]{KrumholzTan2007}. However, 
the relative importance of the above processes for suppressing star formation is unknown, 
even for the case of our own Galaxy. Other basic questions such as ``What is the typical 
lifetime of giant molecular clouds (GMCs)?'', ``What processes initiate star formation in 
localized clumps within GMCs?'', ``Do star-forming clumps come close to achieving virial 
and pressure balance?'' and ``Is the star cluster formation timescale long 
\citep[][]{TanKrumholzMcKee2006} or short \citep[][]{Elmegreen2000,Elmegreen2007} 
compared to free-fall?'' are still debated.

To investigate the star formation process within molecular clouds, a
significant range of the internal structure of GMCs needs to be resolved including dense 
gas clumps expected to be the birth locations of star clusters. In one scenario of GMC 
formation and evolution, large-scale colliding atomic flows have been invoked. 
High-resolution simulations \citep[][among others]{FoliniWalder1998,WalderFolini2000,
KoyamaInutsuka2002, VazquezSemadenietal2003,VazquezSemadenietal2011, VanLooetal2007, 
VanLooetal2010, Hennebelleetal2008, Heitschetal2008, Banerjeeetal2009, AuditHennebelle2010, 
Clarketal2012} show that cold, dense clumps and cores form in the swept-up gas of the 
shock front and that thermal and dynamical instabilities naturally give rise to a 
filamentary and turbulent cloud. 
However, there is little observational evidence that GMCs form from such large-scale and 
rapid converging flows of atomic gas, especially in molecular-rich regions of galaxies, 
such as the Milky Way interior to the solar orbit. For example, even though GMCs are often 
associated with spiral arms in galaxies, in the case of M51, \citet[][]{Kodaetal2009} 
find that the molecular to atomic gas mass fraction does not vary significantly from 
inter-arm to arm regions and infer relatively long GMC lifetimes $\sim 100$~Myr. The 
position angles of projected angular momentum vectors of Galactic GMCs show random 
orientations with respect to Galactic rotation \citep[][]{Kodaetal2006,ImaraBlitz2011}, 
which is inconsistent with young, recently-formed GMCs in the simulations of 
\citet[][]{Tasker&Tan2009}. Similar results are found even in relatively metal and molecular poor 
systems. In the LMC, \citet[][]{Kawamuraetal2009} infer GMC lifetimes of $\sim 26$~Myr, 
significantly longer than their dynamical times. In M33, \citet[][]{Rosolowskyetal2003} 
and \citet[][]{Imaraetal2011} again find random orientations of the position angles of 
projected GMC rotation vectors.

In this paper we investigate the processes of GMC evolution and star formation in a 
kiloparsec-scale patch of a galactic disk, starting from initial conditions in which 
GMCs have already formed. These are extracted from large, global galaxy simulations and 
then the evolution of the interstellar medium (ISM), especially GMCs, and star formation
is followed over a relatively short timescale of $\sim 10$~Myr. This is less than the 
flow crossing time of the simulation volume, but relatively long compared to the dynamical 
and free-fall times of the clouds. We achieve a minimum resolution of $\sim$0.5~pc --- 
enough to begin to resolve a significant range of the internal structure of the GMCs.

This paper, the first of a series, introduces the simulation set up, and then investigates 
the effect of relatively simple physics, including: pure hydrodynamics (no magnetic fields, 
which are deferred to a future paper); a cooling function that approximates the transition 
from atomic to molecular gas and allows cooling all the way down to $\sim 5$~K; 
photoelectric heating; and simple recipes for star formation, parameterized to be a fixed 
formation efficiency per local free-fall time, $\epsilon_{\rm ff}$. Localized feedback 
from newborn stars is difficult to resolve in this simulation set-up, so its treatment is 
deferred to a future paper. Thus the results of these simulations, which include the 
structural and kinematic properties of GMCs and clumps and their star formation rates, 
should be regarded as baseline calculations in a nonmagnetic, zero-feedback
limit. As we shall see, by comparison with observed systems the degree of over production 
of dense gas and stars then informs us on the magnitude of suppression clump 
formation that is needed by these effects.

\section{Methods and Numerical Set-up}\label{sect:Numerical set-up}
\subsection{Simulation Code and Initial Conditions}\label{sect:Local simulations}
The global galaxy simulations of \citet[][hereafter TT09]{Tasker&Tan2009} followed the 
formation and evolution of thousands of GMCs in a Milky-Way-like disk with a flat rotation
curve. However, with a spatial resolution of $\sim$8~pc, only the general, global 
properties of the GMCs could be studied; not their internal structure. Following all of 
these GMCs to higher spatial resolution is very expensive in terms of computational 
resources. Therefore we need to use an alternative method.

\begin{figure*}
\begin{center}
\includegraphics[width=\textwidth]{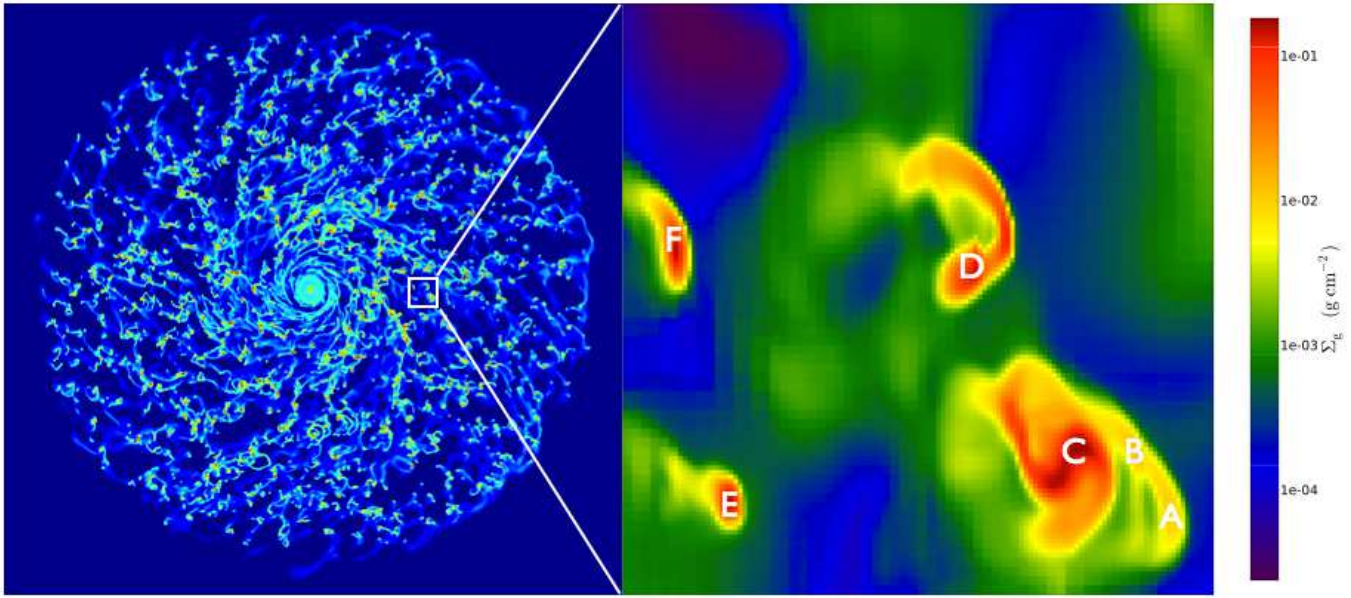}
\caption{View of the gas mass surface density, $\Sigma_g$, over the 20\ kpc diameter galactic 
disk of TT09 250~Myr after start of the simulation. The square is a 1~kpc sided region, 
enlarged in the right image, showing several GMCs. These are the initial conditions for the 
simulations of this paper. The intial clouds listed in Table~\ref{tab:GMCs} are marked.}
\label{fig:region_selection}
\end{center}
\end{figure*}

Our approach is to extract a 1 kpc by 1 kpc patch of the disk, extending 1~kpc both above 
and below the midplane, centered at a radial distance of 4.25~kpc from the galactic center 
(see Fig.~\ref{fig:region_selection}). This is done at a time 250~Myr after the beginning 
of the TT09 simulation, when the disk has fragmented into a relatively stable population 
of GMCs. We then follow the evolution of the ISM, especially the GMCs, including their
interactions, internal dynamics and star formation activity, down to $\lesssim$ 
parsec scales. These local simulations are able to reach higher densities, resolve smaller 
mass scales and include extra physics compared to the global simulations.

Setting up the local simulations requires performing a velocity transformation to remove 
the circular velocity at each location and then adding a shearing velocity field so that 
the reference frame of the simulation is the local standard of rest at the center of the
extracted patch of the galaxy disk. Periodic boundary conditions are introduced for the 
box faces perpendicular to the shear flow (along the $y$-axis) and outflow boundary 
conditions for the other faces. This set up is similar to shearing box simulations of 
astrophysical disks, but does not include the rotation of the frame of the patch (thus 
neglecting Coriolis forces). Since we only consider the evolution of the ISM over quite 
short timescales $\sim 10$~Myr (which is several local dynamical times of the GMCs, but much 
shorter than the orbital time of $130$~Myr), this approximation should not have a significant 
effect on the properties of the clouds and their star formation activity.

We also ignore radial gradients in the galactic potential, only including resulting forces 
perpendicular to the disk. The model for the potential is that used by TT09, i.e. 
\citet[][]{BinneyTremaine1987}, evaluated at $r=4.25$~kpc:
\begin{equation}
\Phi = \frac{1}{2}v_{c,0}^2\ln\left[\frac{1}{r_c^2}\left(r_c^2 + 
r^2 + \frac{z^2}{q_\phi^2}\right)\right],
\end{equation}
where $v_{c,0}$ is the constant circular velocity in the limit of large radii, here set 
equal to $200\:{\rm km\:s^{-1}}$, $r_c$ is the core radius set to 0.5~kpc, $r$ and $z$
are the radial and vertical coordinates respectively, and the axial ratio of the potential 
field is $q_\phi = 0.7$.

The grid resolution of the initial conditions is 128$^2 \times$ 256 which corresponds to a 
cell size of 7.8\ pc and serves as the root grid for the high resolution simulations. Most 
of the simulations we present here involve 4 levels of adaptive mesh refinement of the root
grid, thus increasing the effective resolution to 2048$^2 \times 4096$ or about 0.5~pc. 
Refinement of a cell occurs when the Jeans length drops below four cell widths, in accordance 
with the criteria suggested by \citet[][]{Trueloveetal1997} for resolving gravitational
instabilities.

The simulations performed in this paper were run using {\it Enzo} \citep[][]{BryanNorman1997, 
Bryan1999,OSheaetal2004}. We use the second order Godunov scheme with the Local 
Lax-Friedrichs (LLF) solver and a piecewise-linear reconstruction to evolve the gas equations. 
Because gas temperatures calculated from the total energy can become negative when the total 
energy is dominated by the kinetic energy, we also solve the non-conservative internal energy 
equation.  We use the gas temperature value from the internal energy when the internal energy 
is less than one tenth of the total energy and from the total energy otherwise. While this is 
higher than the often adopted ratio of 10$^{-3}$, it does not affect the dynamics.

\begin{table}
\caption{Star Particle Creation. \label{tab:sf}}
\begin{center}
\begin{tabular}{c c c c c}
\tableline
\tableline
Cell Size & $n_{\rm H,sf}$ & $t_{\rm ff}$ & Min. Cell Mass & $M_{\rm *,min}$\\
(pc) & (${\rm cm^{-3}}$) & (yr) & ($M_\odot$) & ($M_\odot$)\\
\tableline
7.8 & 100 & $4.3\times 10^6$ & 1640 & 1000$^a$ \\
0.49 & $10^5$ & $1.4\times 10^5$ & 400 & 100\\
0.125 & $10^6$ & $4.3\times 10^4$ & 63 & 10$^b$\\
\tableline
\end{tabular}
\end{center}
$^a$ Used in test runs not presented here and by Tasker (2011)\\
$^b$ Used in runs to be presented by Butler, Van Loo \& Tan, in prep.\\
\end{table}

\subsection{Star Formation}
To model star formation, we allow collisionless star cluster particles, i.e. a point mass 
representing a {\it star cluster or sub-cluster} of mass $M_*$, to form in our simulations. 
These star cluster particles are created when the density within a cell exceeds a fiducial 
star formation threshold value of $n_{\rm H,sf} = 10^5$\ cm$^{-3}$ for our 4-level refinement 
runs compared to a threshold of 10$^2$\ cm$^{-3}$ in the lower resolution simulation of 
\citet[][]{Tasker2011}. (Note we assume $n_{\rm He}=0.1 n_{\rm H}$ so that the mass per H is 
$2.34\times 10^{-24}\:{\rm g}$). This density threshold is a free-parameter of our modeling, 
and its choice depends on the minimum mass that is allowed for star particles and the minimum
cell resolution (see Table~\ref{tab:sf}).

We use a prescription of a fixed star formation efficiency per local free-fall time, 
$\epsilon_{\rm ff}$, for those regions with $n_{\rm H}>n_{\rm H,sf}$. Relatively low and 
density-independent values of $\epsilon_{\rm ff}$ are implied by observational studies of GMCs
\citep[][]{ZuckermanEvans1974} and their star-forming clumps \citep[][]{KrumholzTan2007}, which 
motivate our fiducial choice of $\epsilon_{\rm ff} = 0.02$. Such values are also approximately 
consistent with numerical studies of turbulent, self-gravitating gas 
\citep[][]{KrumholzMcKee2005} and turbulent, self-gravitating, magnetized gas 
\citep[][]{PadoanNordlund2011}. Note that given the inability of our simulations to resolve 
individual star-forming cores, we do not impose requirements that the gas flow be converging or 
that the gas structure be gravitationally bound in order for star formation to proceed. However, 
to rule out the possibility of star formation in the hot dense gas of shock fronts, cells with 
temperatures greater than 3000~K are prevented from forming stars. In fact, such conditions 
almost never arise in the simulations presented here.

When a cell reaches the threshold density, a star cluster particle is created whose mass is 
calculated by
\begin{equation}\label{eq:mstar}
 M_* = \epsilon_{\rm ff} \frac{\rho \Delta x^3}{t_{\rm ff}} \Delta t,
\end{equation}
where $\rho$ is the gas density, $\Delta x^3$ the cell volume, $\Delta t$ the numerical time 
step, and $t_{\rm ff}$ the free-fall time of gas in the cell (evaluated as 
$t_{\rm ff} = (3\pi/32G\rho)^{1/2}$).

An additional computational requirement is the minimum star cluster particle mass, 
$M_{\rm *,min}$, introduced to prevent the calculation from becoming prohibitively slow due 
to an extremely large number of low-mass particles. If the calculated $M_*$ is $<M_{\rm *,min}$ 
then a star cluster particle of mass $M_{\rm *,min}$ is created with probability 
$M_*/M_{\rm *,min}$. In fact given the low value of $\epsilon_{\rm ff}$, the fact that 
$\Delta t \ll t_{\rm ff}$ and our adopted values of $M_{\rm *,min}$, we are always in this 
regime of ``stochastic star formation''.

The motions of the star cluster particles are calculated as a collisionless N-body system. 
Note these are not sink particles: there is no gain of mass by gas accretion. They interact 
gravitationally with the gas via a cloud-in-cell mapping of their positions onto the grid to 
produce a discretized density field. Note, however, that gravitational interactions between 
star cluster particles are thus softened to the resolution of the grid. In reality the 
distribution of stellar mass represented by the star cluster particle would be spread out, 
but by amounts that are not set by the local gas density. Thus the structure of star clusters, 
made up of many simulation star cluster particles, is not well-modeled in our simulations and 
we do not present results on the details of the star clusters that form.

\subsection{Thermal processes}\label{sect:thermal processes}
To describe the thermal behaviour of the ISM, we include a net heating rate per unit volume 
given by
\begin{equation}\label{eq:total_heating}
H = n_{\rm H} [\Gamma - n_{\rm H} \Lambda]\  {\rm erg\ cm^{-3}\ s^{-1}},
\end{equation}
where $\Gamma$ is the heating rate and $\Lambda$ the cooling rate.  Below we describe 
several different methods to model heating and cooling.

The thermal processes introduce a new time scale, i.e. the cooling time
$t_{\rm cool} \equiv E_{\rm int}/|H|$ where $E_{\rm int} = p_g/(\gamma -1)$ is the internal energy, 
which is often much shorter than the dynamical time step set by the Courant condition. To 
prevent the cooling from increasing the numerical cost significantly, we set the numerical 
time step to the hydrodynamical time step and sub-cycle the cooling with smaller time steps, 
i.e. $0.1 t_{\rm cool}$ 
until the total cooling time step equals the dynamical time step.
Then the temperature and internal energy are updated explicitly every 
sub-cycle after which the net heating rate and cooling time are recalculated. This 
sub-cycling prevents overcooling and negative pressures by not resolving the cooling time.  
If the hydrodynamical time step is shorter than the cooling time, no sub-cycling is necessary.

\subsubsection{Simple Photoelectric Heating}
FUV radiation is absorbed by dust grains causing electrons to be discharged, which then heat 
the gas. This photoelectric (PE) heating has long been thought to be the dominant heating 
mechanism in the neutral ISM, including the relatively low extinction portions of GMCs 
\citep[][]{Wolfireetal1995}. We include a PE heating term 
\begin{equation}\label{eq:photoelectric}
   \Gamma_{\rm PE} = 1.3 \times 10^{-24} \epsilon G_0\ \  {\rm erg\ s^{-1}},
\end{equation}
where $\epsilon$ is the heating efficiency and $G_0$ the incident FUV field normalized to 
the \citet[][]{Habing1968} estimate for the local ISM value. We set the heating efficiency to its 
maximum value of $\epsilon=$0.05 for neutral grains \citep[][]{BakesTielens1994}. Also, we 
adopt $G_0 = 4$ as a value appropriate for the radial location of the simulated region in 
a Milky Way-type galaxy, in agreement with \citet[][]{Wolfireetal2003}. Note that this estimated 
heating rate is approximate in the sense that it does not follow the attenuation of the
FUV field in the dense gas, nor its local generation from young star clusters.

\begin{figure}
\begin{center}
\includegraphics[width=8cm]{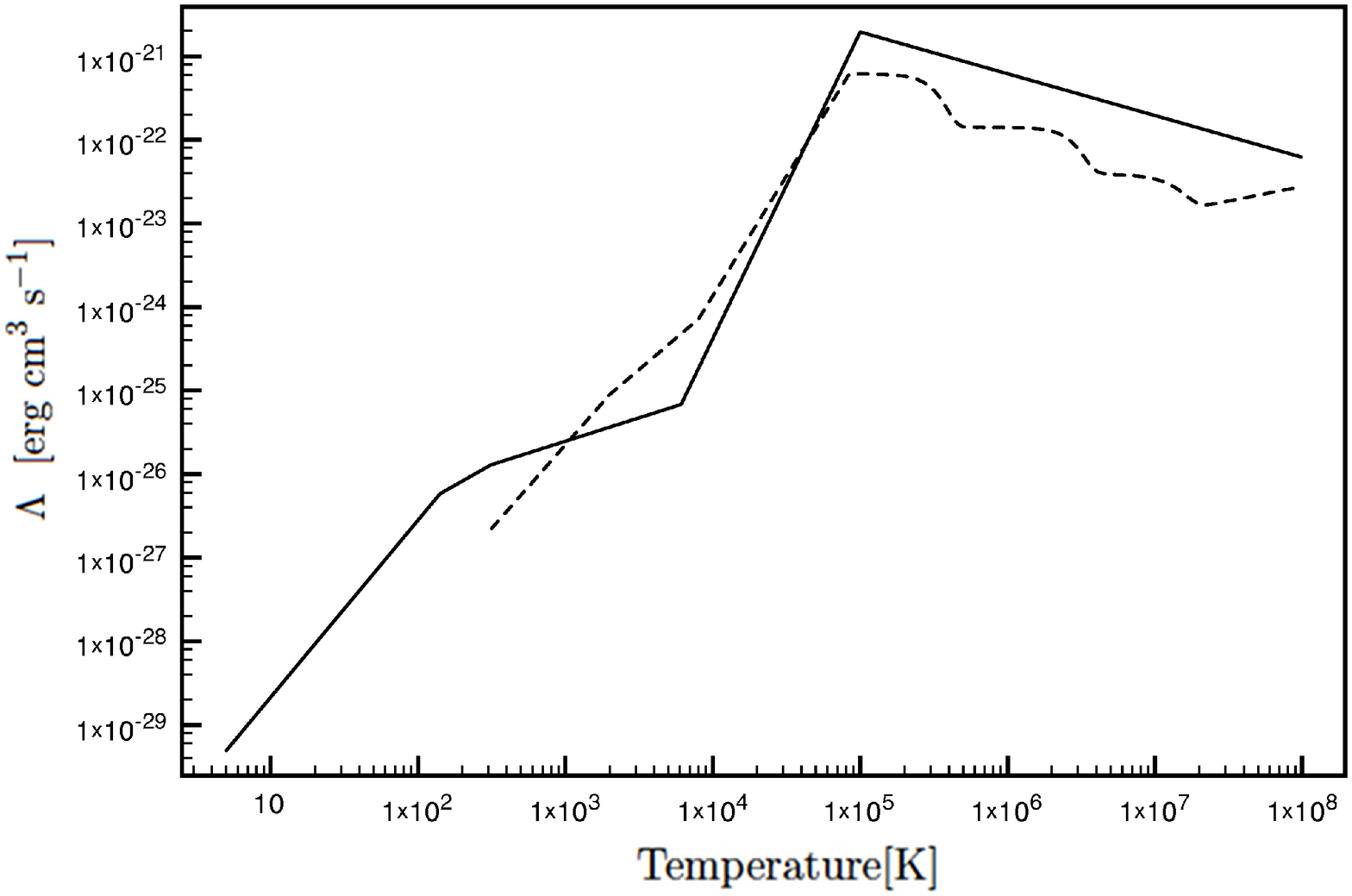}
\includegraphics[width=8cm]{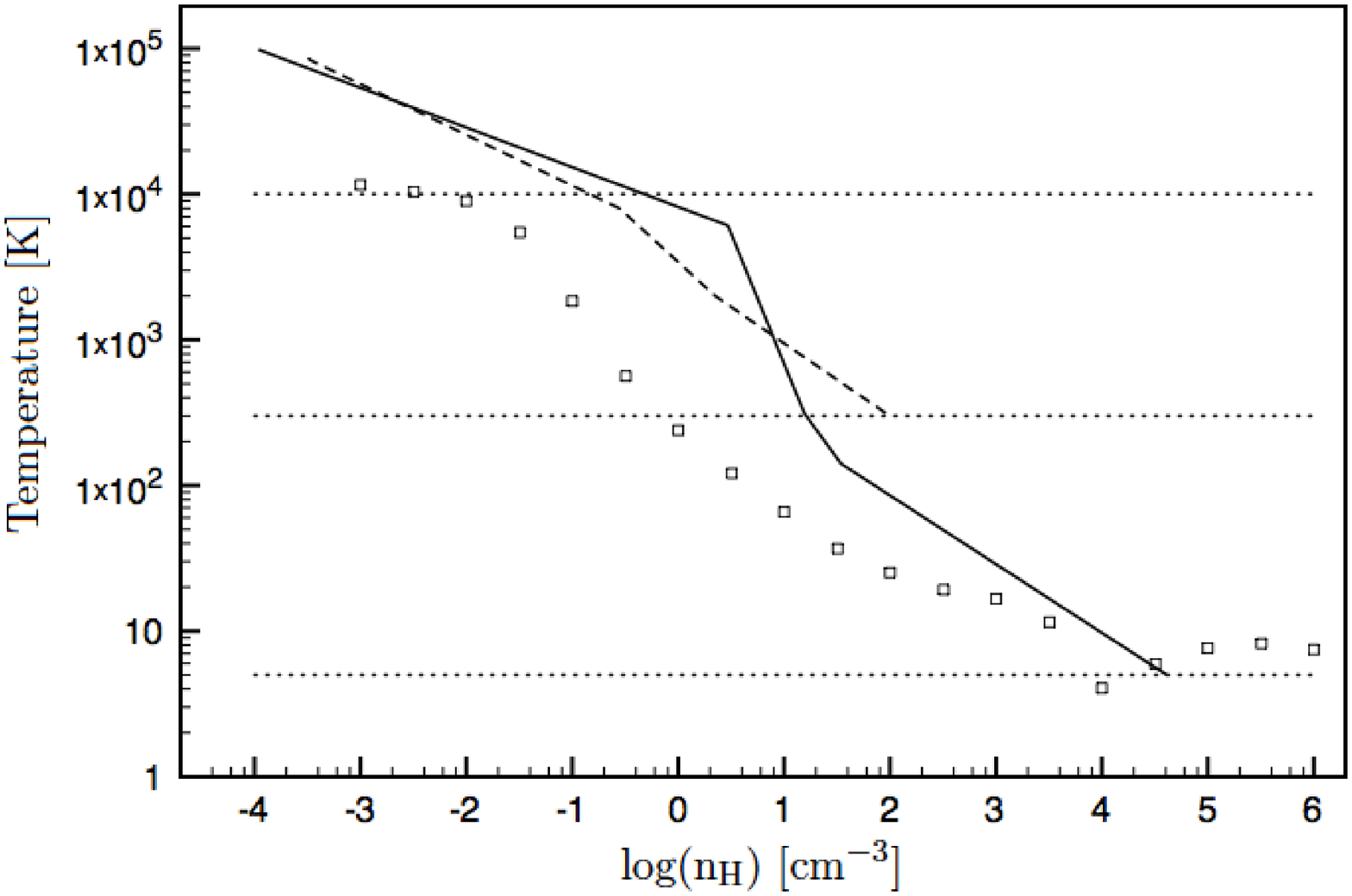}
\caption{{\it Left}: Cooling rate as function of temperature as defined by Sanchez-Salcedo 
et al. (2002) (solid) and Rosen \& Bregman with the high temperature range ($> 10^5$~K) 
from Sarazin \& White (1987) (dashed). {\it Right}: The equilibrium temperatures for 
the Rosen \& Bregman (dashed) and  Sanchez-Salcedo et al. (solid) cooling curves assuming 
a constant PE heating term with $G_0 = 4$ and for the Cloudy heating and cooling rates 
(squares). The dashed lines represent lines of constant temperature at 10$^4$, 300 and 5\ K.}
\label{fig:RBvsSS}
\end{center}
\end{figure}

\subsubsection{Simple Atomic Cooling}
We consider several cooling functions. Initially and for the higher temperature regime, we 
adopt the solar metallicity cooling curve of \citet[][]{SarazinWhite1987} from temperatures 
of $T = 10^8$\ K down to $T = 10^5$\,K  (although we do not expect gas temperatures 
in our simulations to exceed a few times $10^6$\ K) and extend it down to $T = 300$\,K using 
rates from \citet[][]{RosenBregman1995} (see Fig.~\ref{fig:RBvsSS}). These temperatures take us 
to the upper end of the atomic cold neutral medium \citep[][]{Wolfireetal1995}. This cooling 
function is similar to that used by TT09. The floor temperature of 300~K acts as a minimum of 
thermal support against gravitational collapse and fragmentation. It imposes a minimum signal 
speed equal to the sound speed $c_s = (\gamma k T/ \mu m_p)^{1/2} \approx 1.80 
(T/300{\rm K})^{1/2}\ {\rm km\ s^{-1}}$, where $\gamma=5/3$ and $\mu=1.27$ (for an assumed 
$n_{\rm He}=0.1n_{\rm H}$).  This signal speed is of the same order as observed velocity
dispersions of clumps within GMCs \citep[e.g.][]{Barnesetal2011}.

\begin{figure}
\begin{center}
\includegraphics[width=8cm]{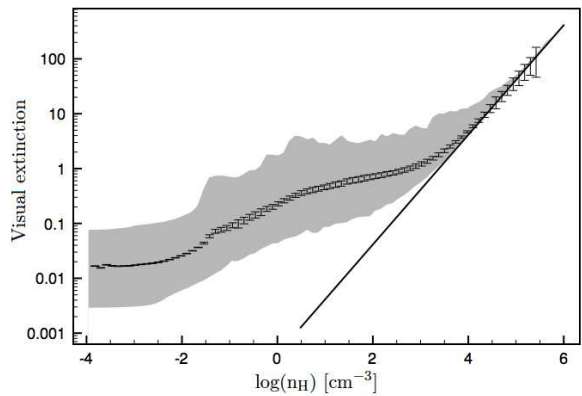}
\caption{Mean logarithmic visual extinction as a function of the density as derived using 
the Sanchez-Salcedo et al. simulation. The error bars show the dispersion on the mean, 
while the shaded area shows the distribution of all column extinctions in the numerical 
domain.  The solid line represents the minimum column extinction due to 
absorption within the cell itself (see Eq.~\ref{eq:selfabsorption}). }
\label{fig:extinction}
\end{center}
\end{figure}

However, the Rosen \& Bregman cooling function does not include the formation and destruction 
of molecules or any cooling processes below the minimum temperature of 300~K. By a combination 
of dust cooling and atomic and molecular line cooling, gas in real GMCs reaches temperatures 
of $\sim 5 - 10$\,K. We first take into account the effect of dust grains by adopting the 
atomic cooling function of \citet[][]{SanchezSalcedoetal2002}, which mimics the equilibrium
phase curve of \citet[][]{Wolfireetal1995}. Figure ~\ref{fig:RBvsSS} shows the difference 
between the Sanchez-Salcedo et al. and the Rosen \& Bregman cooling rates. Note that, while 
the cooling rates are roughly the same for both curves, the Sanchez-Salcedo et al. cooling 
curve extends down to 5\ K and has a thermally unstable temperature range between 313-6102\ K. 
This gives rise to the co-existence of a cold and warm phase at the same pressure.

\subsubsection{Extinction-Dependent Heating and Cooling Functions}\label{sect:molecularcooling}
\paragraph{A density-column extinction relation}
The inclusion of molecular cooling is less straightforward since the formation of molecules 
depends strongly on the amount of attenuation of the radiation field.  Molecules only form 
in the regions of relatively dense gas and dust that can shield their contents from
destructive UV radiation. This means that the molecular cooling rate, and also heating 
rate, depends not only on density and temperature, but also on column extinction.

Of these three variables, the column extinction is the only one that is not directly 
available during the simulation. For every grid cell an effective column 
extinction due to dust absorption can be calculated using a six-ray approximation 
\citep[e.g.][]{Glover+MacLow2007}. Using the linear relation between 
column density and visual extinction, i.e. $A_V = 5.35\times 10^{-22} N_{\rm H}$, the 
effective visual extinction is given by \citep[][]{Glover+MacLow2007}
\begin{equation}
  A_{V,{\rm eff}} = -\frac{1}{2.5}\log 
 	\left[\frac{1}{6} \sum_{i=1}^{6}{\exp(-2.5A_{V,i})}\right] {\rm mag}.
\end{equation}
Although the six-ray approximation already reduces the numerical cost of calculating the 
column extinction considerably, it remains time-consuming as it needs to be calculated 
for every cell at every time step. Still, an additional simplification can be made.  In
general, higher density gas has a higher column extinction than the surrounding lower 
density gas.  We calculate the effective column extinction for the 
high-resolution run with the Sanchez-Salcedo et al. cooling function (see 
Sect.~\ref{sect:SS}) at 10~Myr, but neglect cells within 10 grid spacings of the 
computational boundary. Figure \ref{fig:extinction} shows the full range of column 
extinctions in the numerical domain, i.e. the shaded area, and the mean logarithmic 
column extinction as a function of the density with the error bars indicating the 
dispersion on the mean.  We omitted density bins with fewer than ten cells contributing 
to the mean. While the extrema of the column extinction differ by more than an order of 
magnitude \citep[similar to the results of e.g.][]{Clarketal2012}, the dispersion on 
the mean is much smaller. For example, at $n_{\rm H} = 10\ {\rm cm^{-3}}$, 95\% of 
all visual extinctions lie within the $0.458 \pm 0.021$ interval. Thus, the mean value 
is representative for the column extinction at a given density. An a posteriori check 
on the simulations using this relation (i.e. Run 5, 6 and 7 of 
Table~\ref{tab:sims}) shows little deviation from the above result.

The sudden increase at $n_{\rm H} \approx 10^4 {\rm cm^{-3}}$ is due to the
resolution limitation, i.e. the effective visual extinction is dominated by absorption 
within the cell itself, i.e.  
\begin{equation}\label{eq:selfabsorption}
	A_{V, {\rm eff}} = 5.35\times 10^{-22} \frac{\Delta x}{2} n_{\rm H}~{\rm mag},
\end{equation}
or $A_{V, {\rm eff}} = 4.1~{\rm mag}$ for $n_{\rm H} = 10^4~{\rm cm^{-3}}$ and $\Delta x$ 
= 0.5~pc. The high density end is thus resolution-dependent and the column extinction 
is most likely overestimated. However, for $A_{\rm V} > 10$, the heating rate (and 
ionization balance) is dominated by cosmic rays (see later), so this limitation should 
not affect the cooling and heating rates significantly.

\paragraph{A table of heating and cooling rates}
Using the density-column extinction relation we now generate a table of cooling and 
heating rates as function of density and temperature using the photodissociation code 
Cloudy \citep[][version c08.01]{Ferlandetal1998}.  The routine is similar to the one of
\citet[][]{Smithetal2008} and goes as follows.  Firstly, for a given density, ranging 
from 10$^{-3}$ to 10$^6$ cm$^{-3}$ with steps of 1 dex, the unextinguished local 
interstellar radiation field \citep[][]{Black1987} with $G_0 = 4$ is directed through 
an absorbing slab. This slab has a constant density of $n_{\rm H}=1 {\rm cm}^{-3}$, 
similar to the mean density of the disk in the initial conditions and a thickness
corresponding to the visual extinction at the given density (see 
Fig.~\ref{fig:extinction}).  We assume the gas has solar abundances and include a dust 
grain population with PAHs. The dust grain physics used in the code is described in 
\citet[][]{vanHoofetal2004} and \citet[][]{Weingartneretal2006}. Background cosmic 
rays are also included with a primary ionization rate of 
2.5$\times 10^{-17}\ {\rm s}^{-1}$, as well as the cosmic background radiation.

The transmitted continuum, i.e. the sum of the attenuated incident and diffuse continua 
and lines, is then used as the radiation field incident on a gas parcel with the given 
density. The resulting cooling and heating rates, for a range of temperatures between 5 
and 10$^5$\ K, are calculated self-consistently by simultaneously solving for statistical 
and ionization equilibrium including all the necessary microphysics such as, among others, 
H$_2$ and CO formation and destruction. For temperatures above 10$^5$\ K we opt not to 
use Cloudy and simply adopt the cooling curve of \citet[][]{SarazinWhite1987} and set 
the heating rate to zero. From the generated table, the heating and cooling rates for any 
density and temperature are derived using a bilinear interpolation.  For densities and 
temperatures above and below the table limits, the rate of the limiting value is used. We
implemented such a heating and cooling routine into {\it Enzo}.

\begin{figure}
\begin{center}
\includegraphics[width=8cm]{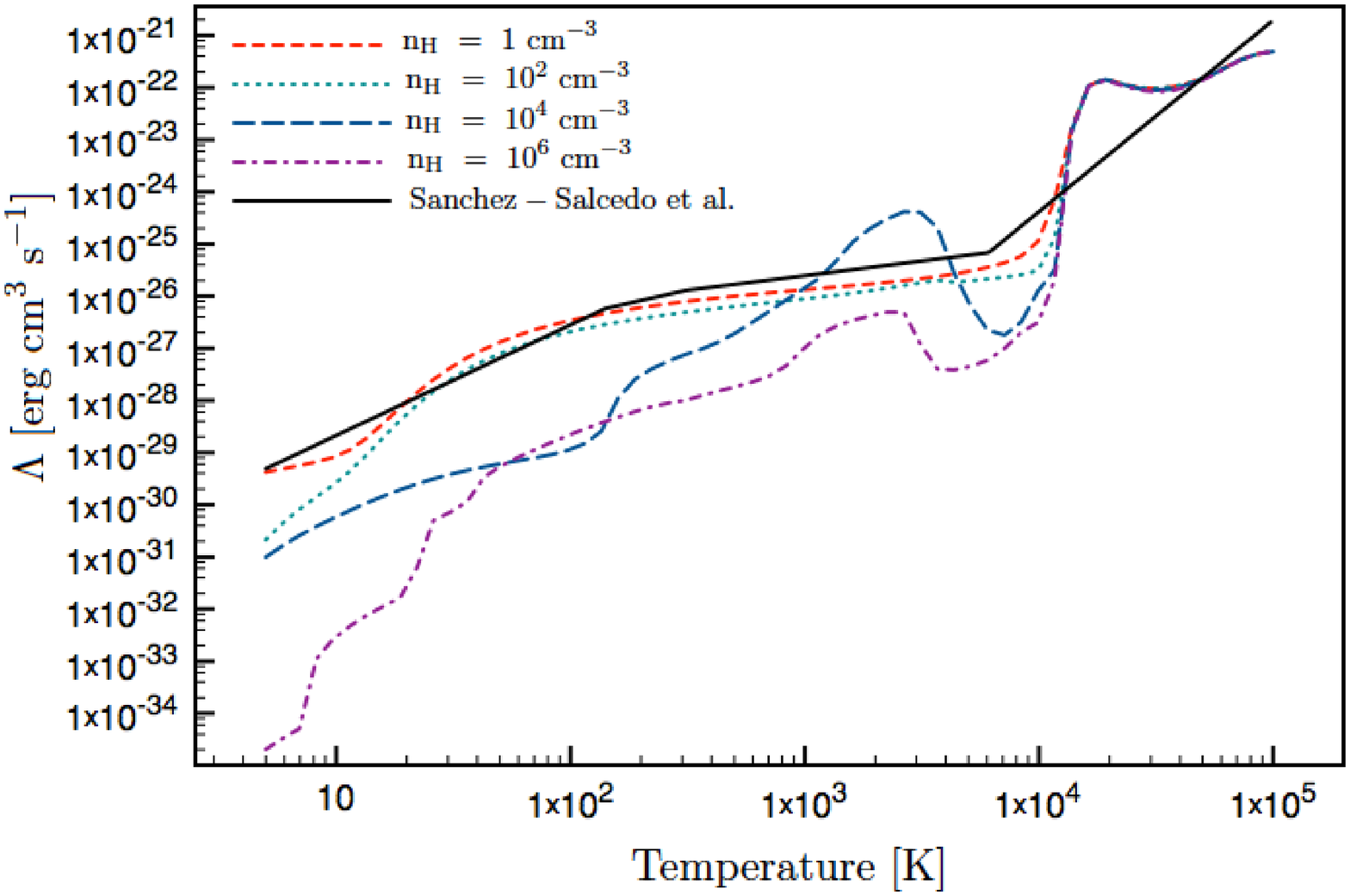}
\includegraphics[width=8cm]{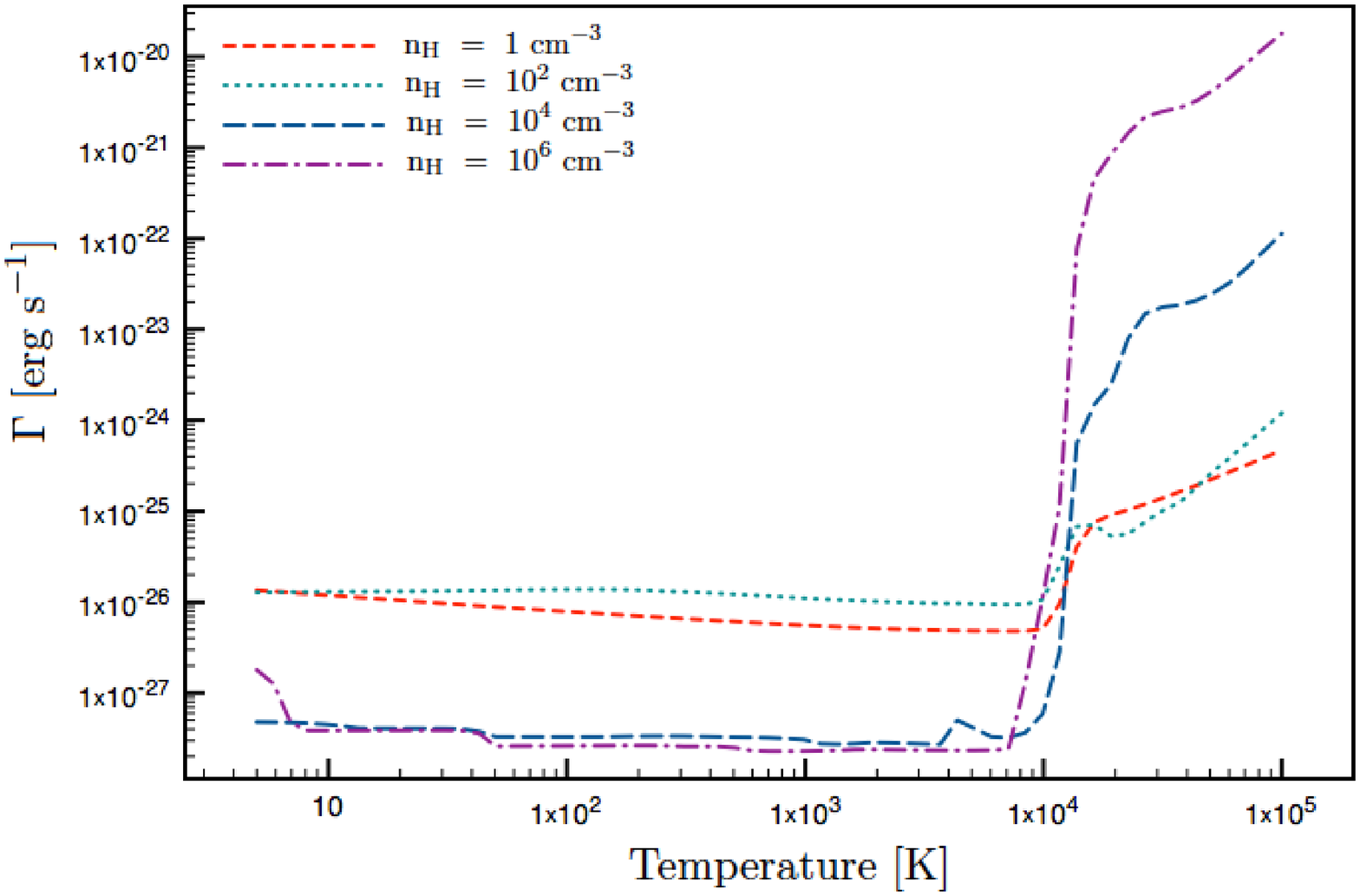}
\caption{The cooling ({\it left}) and heating ({\it right}) rates as function of 
temperature for densities of 10$^0$ (short-dashed), 10$^2$ (dotted), 10$^4$\ cm$^{-3}$ 
(long-dashed) and 10$^6$\ cm$^{-3}$ (dash-dotted) calculated using Cloudy. 
The cooling rates of Sanchez-Salcedo et al. (solid) are also plotted as reference. }
\label{fig:heating_cooling}
\end{center}
\end{figure}

\begin{figure}
\begin{center}
\includegraphics[width=8cm]{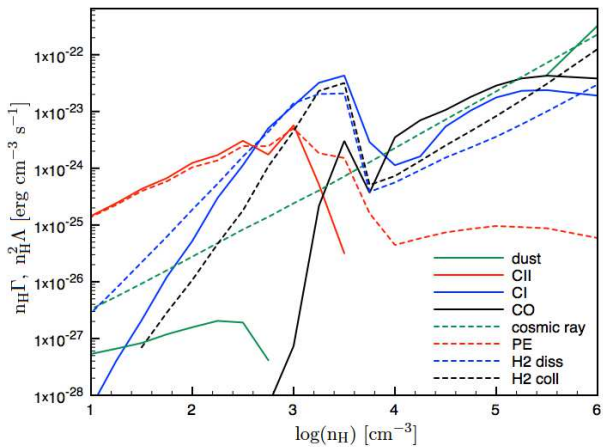}
\caption{Cooling (solid) and heating (dashed) rates per unit volume as 
function of density for the Cloudy cooling function equilibrium temperatures 
given in Fig.~\ref{fig:RBvsSS}.  Only the processes that contribute the most are shown.}
\label{fig:equilibriumcomponents}
\end{center}
\end{figure}

Figure \ref{fig:heating_cooling} shows the calculated heating and cooling rates as a 
function of temperature for selected densities. The Sanchez-Salcedo et al. cooling curve 
is also plotted showing a close similarity in shape and magnitude for densities up to 
$\sim 10^2$~cm$^{-3}$. The deviation from the Sanchez-Salcedo et al. curve at higher 
densities stems from the onset of molecular cooling.  Molecules form in regions with 
visual extinctions higher than $A_V = 2.4$ or column extinctions higher than 
4.3 $\times 10^{21}$ cm$^{-2}$ \citep[e.g.][]{TielensHollenbach1985}. Using the derived
density-column extinction relation this translates to densities above 
$\approx 10^3~{\rm cm^{-3}}$. Figure~\ref{fig:equilibriumcomponents} shows the decomposition 
of the heating and cooling rates at thermal equilibrium (see Fig.~\ref{fig:RBvsSS} for 
the equilibrium temperatures).  From this figure, we indeed find that molecular species, 
i.e. H$_2$ and CO, contribute significantly to the heating and cooling above 10$^3$~cm$^{-3}$.  
Note that this figure is very similar to Fig.~8 of \citet[][]{Glover&Clark2012} in both 
the decomposition and the level of cooling rates. Note also that, the dust cooling 
disappears between $\approx 10^3-10^5$~cm$^{-3}$. Up to 10$^5$~cm$^{-3}$, the dust grains 
are hotter than the gas. In these conditions, ion-grain collisions heat the gas due to 
thermal evaporation of neutralized ions (see Eq. 32 of \citet{Baldwinetal1991}), while 
electron-grain collisions cool the gas.  At densities below 10$^3$~cm$^{-3}$ the 
electron-grain collisions dominate the ion-grain collisions so that the net effect is gas 
cooling.  As the density (and extinction) increases, less ionizing radiation can penetrate 
the gas. Then ion-grain collisions start to dominate the electron-grain collisions and, 
consequently, the gas is heated. Above 10$^5$~cm$^{-3}$ the dust temperature falls below 
the gas temperature and grain collisions result in cooling of the gas.

While the Cloudy cooling rates are adequately described by the Sanchez-Salcedo cooling 
function (up to 10$^2$~cm$^{-3}$), the Cloudy heating rates deviate significantly from 
the PE heating rate we use with the atomic cooling functions, i.e.  $\approx 2.6 
\times 10^{-25}$~erg~s$^{-1}$. It is clear that we have overestimated the heating 
efficiency by approximately an order of magnitude.  At high densities, the PE heating 
is no longer the dominant heating process (see Fig.~\ref{fig:equilibriumcomponents}).
Higher density regions tend to have higher dust extinctions of the external radiation 
field, thus blocking FUV penetration and PE heating. Then only cosmic rays can penetrate 
and heat the gas. The overall heating rate is further reduced by an order of magnitude.

Not only does the decomposition of the cooling and heating rates give useful insights 
into the dominant processes at different temperatures and densities, it also provides 
the opportunity to compare the numerical simulations with observations.  For different 
emission lines, such as the 158 $\mu$m CII line and the 63 $\mu$m OI, tables similar 
to the cooling and heating rate tables can be constructed. Then emissivity maps and line 
profiles of optically-thin emission lines can be generated from the simulations during
post-processing.  Note, however, that such emission maps are only first 
order approximations as they do not include any radiative transfer, nor do they take into
account the effects of time-dependent chemistry \citep[e.g.][]{Glover+MacLow2007,
Glover+MacLow2011} or local abundance variations \citep[][]{Shettyetal2011a,
Shettyetal2011b}.

\paragraph{Thermal instability?}
Although the Cloudy cooling curve (for $n_{\rm H} < 10^2$~cm$^{-3}$) has a similar 
temperature dependence between 300-6000\ K as the Sanchez-Salcedo et al. curve, the 
equilibrium curve only exhibits a weak thermal instability (see Fig.~\ref{fig:RBvsSS}).  
Furthermore, the instability is at lower densities than the Sanchez-Salcedo et al.
curve. The shift of the instability towards lower densities is due to a higher column 
extinction compared to the 10$^{19}$\ cm$^{-2}$ used by \citet[][]{Wolfireetal1995}.  
At low densities, the thermal equilibrium is set by the Ly$\alpha$ cooling and PE 
heating. The increased attenuation reduces the electron fraction in the gas resulting 
in a lower PE heating (and thus a lower equilibrium temperature).

Once the equilibrium temperature drops below 10$^4$\ K (and this happens at lower 
densities), CII cooling becomes the dominant cooling process.  \citet[][]{Wolfireetal1995} 
show that the thermal instability can be attributed to CII cooling which causes a rapid 
drop in equilibrium temperature.  The magnitude and presence of the thermal instability 
then depends on the temperature and density dependence of the CII cooling. Interstellar gas 
is thermally unstable if \citep[][]{Field1965}
\begin{equation}
   \frac{n_{\rm H}}{T} \left(\frac{\partial H}{\partial n_{\rm H}}\right)_T - 
        \left(\frac{\partial H}{\partial T}\right)_{n_{\rm H}} < 0.
\end{equation}
By expressing the heating and cooling rate locally as a power-law of density and
temperature, the instability constraint reduces to 
\begin{equation}\label{eq:thermalinstability}
        (a - b) - (\alpha - \beta) < 1,
\end{equation}
where $a$, resp. $b$, is the density, resp. temperature, power-law index for the heating 
rate and $\alpha$ and $\beta$ the indices for the cooling rate.  We can use the above 
constraint to understand the weak instability seen in Fig.~\ref{fig:RBvsSS}. For 
temperatures between $300 - 6000$\ K the cooling rate has a power-law index of the order 
of 0.5, while the heating rate shows no temperature dependence, i.e. $\beta \approx 0.5$ 
and $b \approx 0$ (see Fig.~\ref{fig:heating_cooling}). Similarly, while the heating rate 
index, $a$, is roughly zero between 0.1 and 1 cm$^{-3}$ (the range of densities for which 
the equilibrium temperatures is between 300 and 6000\ K), the cooling rate decreases 
roughly as $n_{\rm H}^{-0.5}$. As a result, $(a - b) - (\alpha - \beta) \approx 1$.  The 
instability criterion is thus only marginally satisfied explaining the weak thermal 
instability.

\begin{table}
\caption{Set of simulations. \label{tab:sims}}
\begin{center}
\begin{tabular}{c c c c c c}
\tableline
\tableline
Run & AMR$^a$ & Heating & Cooling & $\mu$ & Star formation \\
\tableline
1 & No & No & RB$^b$ & 1.27 & No\\
2 & Yes & No & RB & 1.27 & No\\
3 & Yes & PE & RB & 1.27 & No\\
4 & Yes & PE & SS$^c$ & 1.27 & No\\
5 & Yes & Cloudy& Cloudy & 1.27 & No \\
6 & Yes & Cloudy& Cloudy & 2.33 & No \\
7 & Yes & Cloudy& Cloudy & 2.33 & Yes \\
\tableline
\end{tabular}
\end{center}
 $^a$ ``No'' implies min. resolution of $8$~pc; ``Yes'' implies min. resolution of 0.5~pc\\
 $^b$ Rosen \& Bregman cooling function\\
 $^c$ Sanchez-Salcedo et al. cooling function\\
 $^d$ Photoelectric heating
\end{table}
 
\section{Results}
To understand the effect of different physical processes, we gradually include them 
in our models.  All the details (e.g. included physics) of the simulations that we 
ran are listed in Table~\ref{tab:sims}.

\subsection{Measures of ISM Structure}\label{sect:measures}
Several methods can be used to compare the different simulations and quantify the 
effect of the included physical processes. We focus on measures that describe the 
changes in density and temperature.

A visual analysis of the simulations is the simplest and quickest method of studying 
the geometrical changes. Column density plots show how the density structures change 
over time, with the maximum/minimum values indicative of increases/decreases in the 
density. 

We also follow the evolution of the mass fraction of the gas that is in defined density 
ranges: ``GMC'' gas is defined by $n_{\rm H} > 10^2$\ cm$^{-3}$  and 
``Clump'' gas is defined by $n_{\rm H} > 10^5$\ cm$^{-3}$.  Using our 
density-column extinction relation, this corresponds to a mean $A_{V,\rm eff} > 0.7$ for 
GMC gas and $A_{V, \rm eff} > 40$ for clump gas. While a visual extinction of 0.7 does 
not imply molecular gas, it is only above $n_{\rm H} = 10^2 {\rm cm^{-3}}$ that we find 
gas with $A_{V, {\rm eff}} > 2.4$ (see Fig.~\ref{fig:extinction}). The density and 
column density structure can also be studied in more detail by using mass weighted 
probability density functions at different times.

As the ISM has different phases, we also examine the mass fractions of the different 
temperature regimes, i.e. the cold, unstable, warm and hot phases.  For the cold phase, 
$T < 350$\ K. The unstable phase lies within the range of 350-6000\ K. The warm phase 
extends from 6000 to 10$^5$\ K after which the hot phase starts. These values are only
indicative and we can even argue whether we need to include a thermally unstable range. 
Technically the Rosen \& Bregman cooling function does not have one and the thermal 
instability in the Cloudy cooling function is weak.

Finally, we also use a clump-finding routine \citep[][for details]{Smithetal2009} to 
identify individual molecular clouds and derive their properties such as their mass, 
velocity dispersion and size. For this purpose, we adapted the clump-finding routine 
included in YT \citep[][]{Turketal2010}. This routine can also be used to identify 
dense clumps within the molecular clouds. However, as the highest resolution in our 
simulations is only 0.5 pc, clumps with sizes of the order of 1\ pc are not properly 
resolved. Therefore, we refrain from any detailed analysis of the clumps.

\begin{table}
\caption{List of the initial GMCs. \label{tab:GMCs}}
\begin{center}
\begin{tabular}{c c c c c c}
\tableline
\tableline
Cloud & Mass center & Mass &  $\sigma^a$ & R$^{b}$ & $\alpha_{\rm vir}^{c}$\\
 &  position (x,y,z) & (10$^6$ M$_\odot$)  & (km\ s$^{-1}$)& (pc) & \\
\tableline
A & 0.921, 0.164, -0.004&0.09& 3.6&17.9 & 1.70\\
B & 0.865, 0.249, 0.004&0.06&  2.5&15.1 & 1.50\\
C & 0.745, 0.232, 0.013&7.47 & 15.4& 47.9 & 0.61\\
D & 0.605, 0.597, 0.005&2.53 & 16.1& 37.4 & 1.54\\
E & 0.180, 0.159, -0.003&0.79 & 10.5& 21.5 & 1.26\\
F & 0.089, 0.588, -0.003&1.00 & 10.8& 27.0 & 1.32\\
\tableline
\end{tabular}
\end{center}
$^{a}$ $\sigma$ is the mass-weighted three-dimensional velocity dispersion.\\
$^{b}$ The radius is calculated from the cloud's volume assuming a spherical geometry.\\
$^{c}$ The virial parameter is calculated as $\alpha_{\rm vir} = \frac{5\sigma_c^2 R}{GM}$, where
$\sigma_c$ is the one-dimensional velocity dispersion inside the cloud and given by
$\sigma_c^2 = \frac{\sigma^2}{3} + c^2_s$ \citep[][]{Dibetal2007}.
\end{table}

\subsection{Initial Conditions}\label{sect:initialconditions}
Figure~\ref{fig:region_selection} shows the column density integrated along the $z$-axis 
and we can identify four distinct density structures. In fact, the selected region contains
six clouds. The two small clouds (A \& B) are about to interact with a larger
one (C) and are therefore not recognized as individual clouds in the column density plot. 
Table~\ref{tab:GMCs} lists all of the clouds. Their properties span a large range in sizes 
and masses.  The smallest cloud only has a radius of 15\ pc and a mass of 6$\times 10^4
M_\odot$, while the largest cloud is a hundred times more massive 
($7.47 \times 10^6 M_\odot$) and 30 times larger in volume. The total mass in the clouds 
is 1.2$\times 10^7 M_\odot$ which is about 70\% of the gas mass in the simulation box.  
The clouds have diameters smaller than $\sim$ 100~pc, and thus have at most $\sim 12$ 
cells ($\Delta x \approx 7.8$\ pc) across each linear dimension.  
To resolve turbulence in self-gravitating gas, \citet[][]{Federrathetal2011}
suggest that the turbulent length-scale needs to be resolved by at least 30 grid cells.
The internal structure of the clouds and their internal turbulence is thus not resolved 
in the initial conditions.  Note, however, that the velocity dispersion of the clouds is 
significantly larger than the minimum sound speed of 1.8\ km\ s$^{-1}$ and increases
with the cloud radius.  Actually, the  initial velocity dispersion of the 
clouds is high enough to give some approximate balance against self-gravity. The mean 
virial parameter of the clouds is 1.32 with standard deviation of 0.35, close to the 
value of 1.3 with standard deviation of 0.76 for Galactic GMCs derived by 
\citet[][]{McKeeTan2003} from analysis of the results of \citet[][]{Solomonetal1987}. 
Note the $\rm ^{13}CO$-selected clouds studied by \citet[][]{Romanduvaletal2010}, which 
trace somewhat higher densities, have median virial parameters of 0.46. The values listed 
here are also in agreement with the values from the simulations 
of \citet[][]{DobbsBurkertPringle2011}.

The temperature distribution of the initial conditions shows that 99\% of the mass has 
a temperature below 350~K and is therefore in the cold phase.  All this gas lies within 
the galactic disk. The remaining 1\% of the mass is diffuse hot gas with temperatures 
above 10$^5$\ K surrounding the disk. TT09 were not studying the full temperature structure 
of the warm gas and so did not yet include a heating term in their simulations. Thus 
most of the gas cooled down to the minimum temperature of 300~K, with hotter components 
created in shocks. The cooling time for low density gas with temperatures above 10$^5$\ K 
(i.e. the gas outside the disk) is of the order of 10$^4$\ Myr and, thus, remains hot. 
As the hot gas lies outside the disk, its volume fraction can be used to derive a mean
thickness of the disk, i.e. 140\ pc.

\subsection{Evolution of Structure and Effect of Resolution}\label{sect:higher}
We first carryout the simulation, running for 10~Myr, with the physics and resolution 
identical to the global galaxy simulation of TT09. 

For the given resolution of $\sim$ 7.8 pc and the minimum temperature of 300\ K, 
the \citet[][]{Trueloveetal1997} criterion, i.e. the Jeans length, 
$\lambda_J = (\pi c_s^2/G\rho)^{1/2}$ where $G$ is the gravitational constant, 
associated with a grid cell should be at least 4 times larger than the size of the 
cell, $\Delta x$, is satisfied for densities up to $n_{\rm H}\simeq 80\:{\rm cm^{-3}}$. 
Similar considerations for the higher resolution simulations that include cooling down 
to $\sim 5$~K and reach much higher densities indicate that the Truelove criterion is 
not satisfied for resolving clump formation within GMCs. Artificial fragmentation is 
expected to happen above densities of 5$\times 10^2$, 1.5$\times 10^3$ and 
2$\times 10^4$~cm$^{-3}$ for the Cloudy, Sanchez-Salcedo et al. and Rosen \& Bregman 
cooling functions, respectively.

\begin{figure}
\begin{center}
\includegraphics[width=8cm]{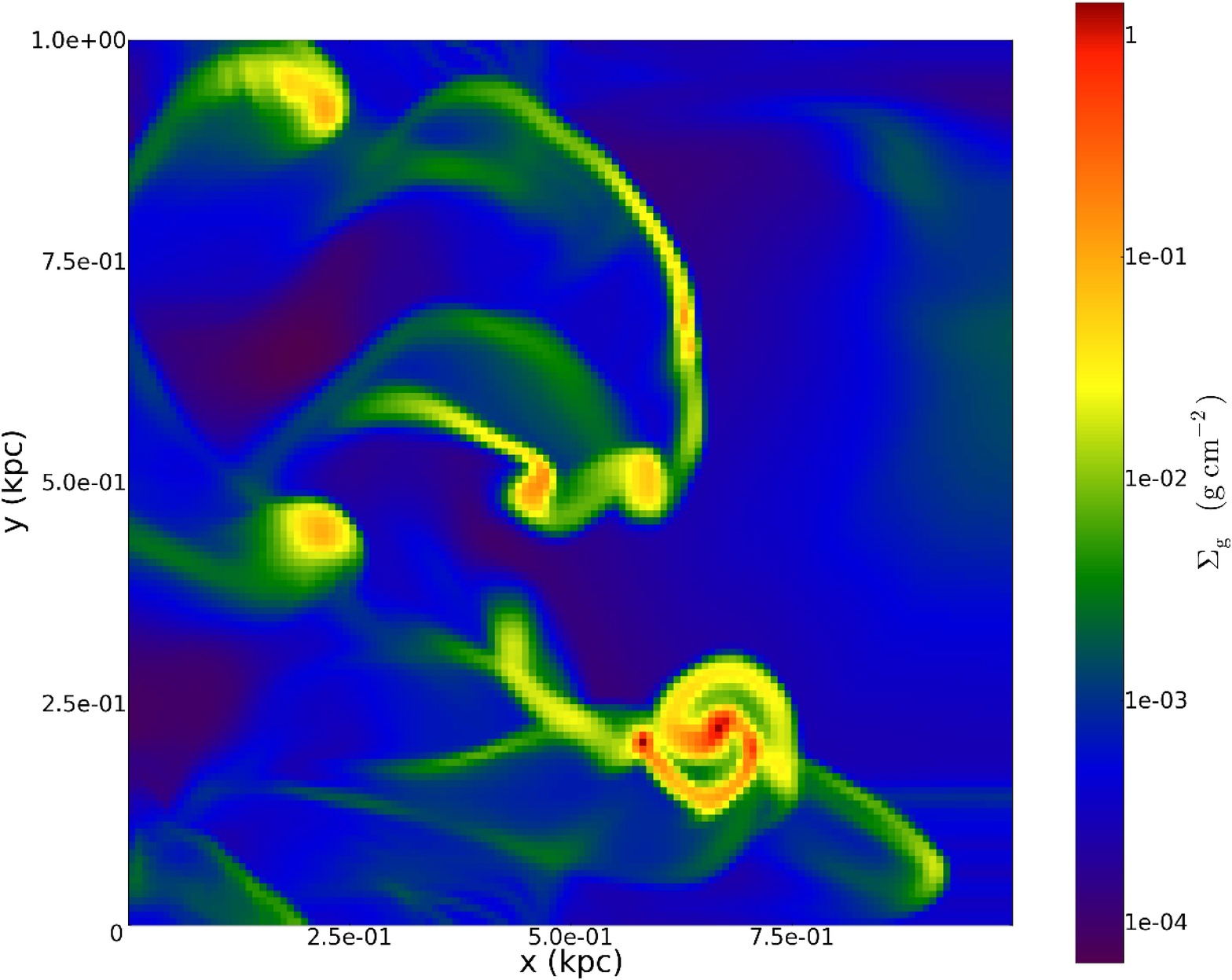}
\includegraphics[width=8cm]{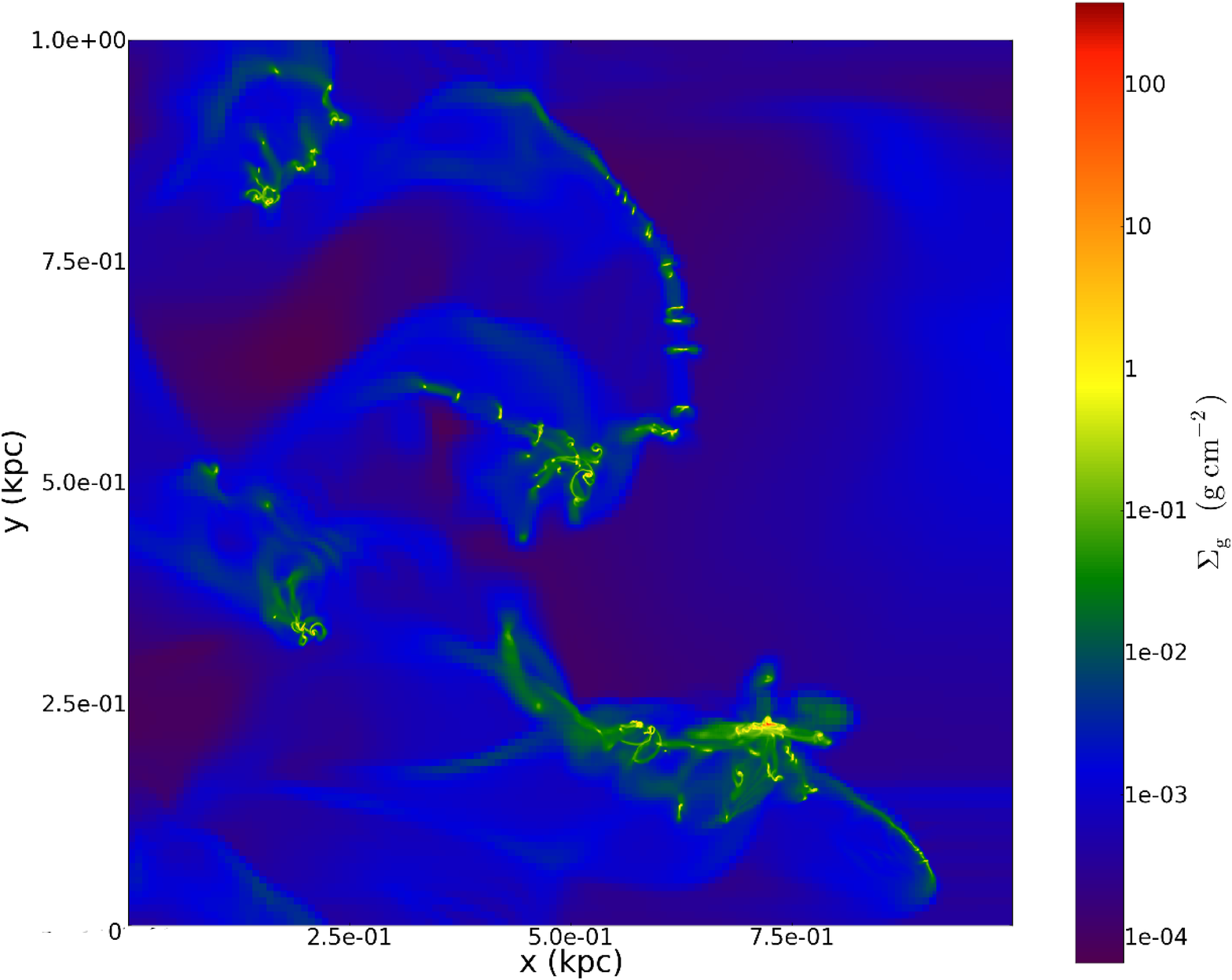}
\caption{Gas mass surface density along the $z$-axis after 10\ Myr for the uniform
Run 1 (left) and the AMR Run 2 (right).}
\label{fig:uni_amr}
\end{center}
\end{figure}

\begin{figure}
\begin{center}
\includegraphics[width=8cm]{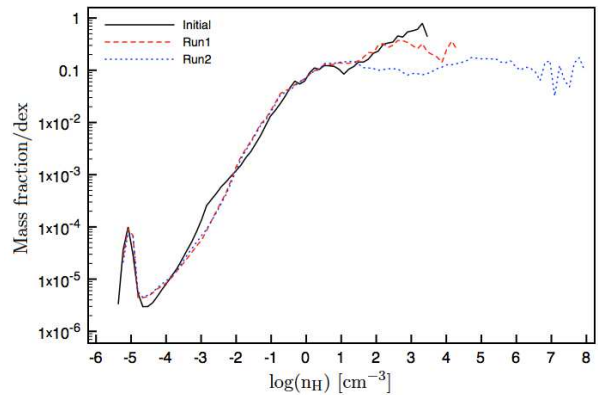}
\caption{Mass-weighted Probability Density Function (PDF) for Run 1 (red, dashed) and 
Run 2 (blue, dotted) after 10\ Myr. We also plot the initial PDF (solid line).}
\label{fig:PDF_hires}
\end{center}
\end{figure}

\begin{figure}
\begin{center}
\includegraphics[width=8cm]{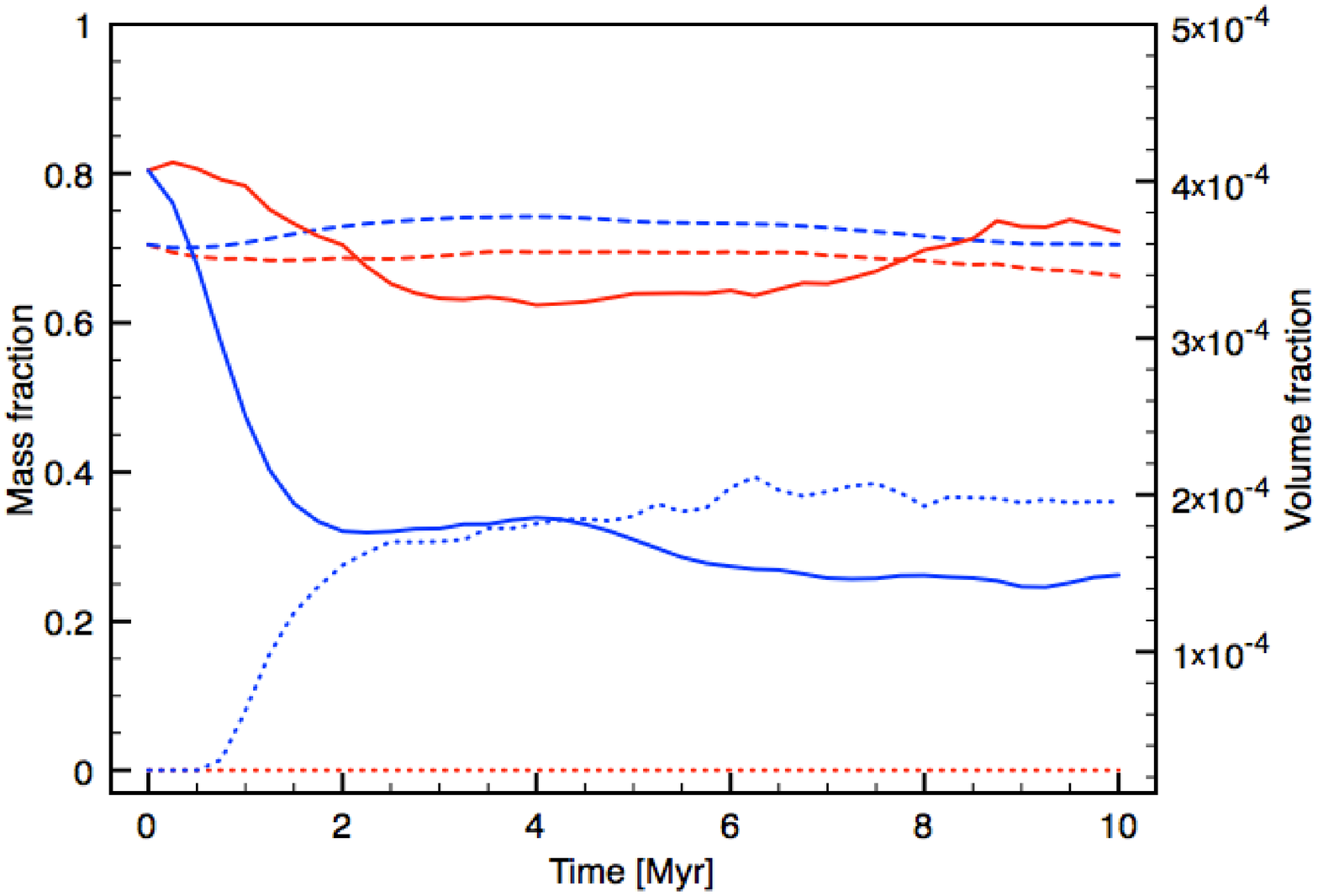}
\caption{The mass fraction of gas in ``GMCs'' ($n_{\rm H} > 10^2 {\rm cm^{-3}}$; dashed)
and in ``Clumps'' ($n_{\rm H} > 10^5 {\rm cm^{-3}}$; dotted) and the volume fraction of 
gas in ``GMCs'' (solid) for Run 1 (red) and Run 2 (blue).}
\label{fig:molecular_mfrac}
\end{center}
\end{figure}

\begin{table}
\caption{List of GMCs after 10 Myr for Run 1 and 2. \label{tab:endGMCs}}
\begin{center}
\begin{tabular}{c c c c c}
\tableline
\tableline
Cloud &  Mass &  $\sigma$ & R & $\alpha_{\rm vir}$\\
Run 1 &   (10$^6$ M$_\odot$)  & (km\ s$^{-1}$)& (pc) & \\
\tableline
A$^a$ & - & - & - & -\\
B$^a$ & - & - & - & -\\
C & 7.7 & 17.7& 44.7 & 1.04\\
D1$^b$ & 0.41 & 5.1 & 22.3 & 1.10\\
D2$^b$ & 0.51 & 4.8 & 25.5 & 0.73\\
D3$^b$ & 0.87 & 6.9 & 25.3 & 0.92\\
E & 0.70 & 5.5 & 26.1 & 0.81\\
F & 0.76 & 6.7 & 27.6 & 0.99\\
\tableline
\end{tabular}
\end{center}
\begin{center}
\begin{tabular}{c c c c c}
\tableline
\tableline
Cloud &  Mass &  $\sigma$ & R & $\alpha_{\rm vir}$\\
Run 2 &   (10$^6$ M$_\odot$)  & (km\ s$^{-1}$)& (pc) & \\
\tableline
A$^a$ & - & - & - & -\\
B$^a$ & - & - & - & -\\
C1$^b$ & 1.1 & 13.6& 21.9 & 2.2 \\
C2$^b$ & 6.9 & 25.6& 29.0 & 4.1 \\
D1$^b$ & 0.10 & 6.4 & 9.3  & 3.1 \\
D2$^b$ & 0.12 & 5.5 & 10.2 & 2.0 \\
D3$^b$ & 0.48 & 16.4& 12.3 & 5.4 \\
D4$^b$ & 0.12 & 14.8& 18.0 & 1.9 \\
D5$^b$ & 0.19 & 13.0& 6.4  & 4.5 \\
D6$^b$ & 0.11 & 8.4 & 6.4  & 2.9 \\
E & 0.70 & 15.2& 14.7 & 2.9 \\
F1$^b$ & 0.16 & 6.8 & 9.6  & 1.6 \\
F2$^b$ & 0.81 & 12.4& 15.4 & 2.0 \\
\tableline
\end{tabular}
\end{center}
$^a$ Cloud A and B merge with cloud C.\\
$^b$ The cloud fragments into multiple clouds. 
\end{table}

Figure~\ref{fig:uni_amr} shows the gas surface density for Run 1 (left panel) after 
10\ Myr. The clouds have not changed dramatically over this time scale, but do show 
signs of gravitational contraction, interaction (e.g. clouds A \& B merge and collide 
with cloud C) and fragmentation (e.g. cloud D).   The result of these 
interactions can be seen in the mass weighted PDFs (Fig.~\ref{fig:PDF_hires}), i.e. 
the ``GMC" gas is redistributed over a slightly larger density range. However, the 
same PDF also shows that the mass fraction of gas in ``GMCs" is roughly the same as 
initially. In fact, the fraction remains quite constant throughout the simulation, 
which spans a few free-fall times (Fig.~\ref{fig:molecular_mfrac}). The volume fraction 
of ``GMC" gas shows an initial decrease, but also remains relatively constant. Thus the 
clouds are roughly in virial equilibrium as indicated by their initial and final virial 
parameters (see Tables~\ref{tab:GMCs} and \ref{tab:endGMCs}). Thermal pressure and 
non-thermal motions within the clouds thus provide enough support against self-gravity to 
prevent runaway gravitational collapse (even though the non-thermal motions are not 
well resolved in this uniform grid simulation). A similar observation was made by 
TT09 in their global simulation, i.e. the gas properties in the full disk are quasi-steady 
after timescales of about 150\ Myr.  However, note that now the velocity frame of the 
simulation volume is the local standard of rest of the disk at the center of the box, 
so the fast, $\sim 200\:{\rm km\:s^{-1}}$ orbital velocities that were present in the 
TT09 simulation are now absent. Modeling these fast circular velocities on a finite 
rectilinear grid led to relatively large numerical viscous heating that helped stabilize 
the TT09 GMCs. Thus it is not surprising that we now see that the clouds are able to 
evolve to somewhat higher densities than were seen by TT09.

By increasing the resolution up to $\sim 0.5$\ pc the GMCs can now be 
better resolved and the evolution of dense clumps within the clouds 
begins to be captured. While the ``GMC'' mass fraction increases slightly to 72\%,
the volume fraction decreases by a factor of 2 within 2\ Myr after which it remains 
roughly constant (see Fig.~\ref{fig:molecular_mfrac}). In approximately the same time 
span, about half of the ``GMC'' gas accumulates in ``Clumps'' (see 
Fig.~\ref{fig:molecular_mfrac}).  Then the mass fraction in clumps remains nearly constant 
and reaches a new quasi-steady state. So, initially, the gas within the clouds collapses
to form filaments with dense clumps due to the increased resolution. The clouds thus 
contract, but, at the same time, the velocity dispersion in the clouds increases. Thermal 
pressure and non-thermal motions again counter the effects of self-gravity to virialize 
the cloud (the virial parameters of the clouds are given in Table~\ref{tab:endGMCs}). 
The initial evolution of the clouds is thus a direct consequence of the increase in 
resolution. 

The formation of clumps can also be seen in the PDFs (see Fig.~\ref{fig:PDF_hires}). 
While the PDFs for Run 1 and 2 are similar up to $n_{\rm H} = 80~{\rm cm}^{-3}$, 
the gas above the ``GMC'' density threshold is redistributed towards higher densities.
The higher densities also give rise to higher surface densities, i.e. the maximum
value increases to $\sim 10^2~{\rm g~cm^{-2}}$ (see Fig.~\ref{fig:uni_amr}).

\subsection{A Multiphase Interstellar Medium}\label{sect:SS}
While the increased resolution helps to describe the substructures of GMCs in greater 
detail, the thermal properties of the ISM are poorly reproduced. As only cooling is 
included in Run 1 and 2, most of the gas within the disk is at the floor temperature of 
300\ K. To reproduce the multiphase character of the ISM, i.e. a cold, dense and a 
warm, diffuse phase, we include diffuse heating. Additionally, we study the influence 
of different cooling functions.  While atomic cooling can be adequately described by 
the Rosen \& Bregman cooling function (used in Sect.~\ref{sect:higher}), the 
Sanchez-Salcedo et al. function extends down to a temperature of 5\ K, i.e.  nearly two
orders of magnitude lower, and includes a thermally unstable temperature range.  
Extinction-dependent cooling, especially from CO molecules, is only taken into account 
in the Cloudy cooling function.

Including diffuse heating mostly affects the gas outside the GMCs. Figure~\ref{fig:RBvsSS} 
shows that, for $n_{\rm H} < 90$\ cm$^{-3}$, the gas has a higher equilibrium temperature 
than 300\ K. For the other cooling functions the critical density for heating is at lower 
densities, i.e. 10\ cm$^{-3}$ for Sanchez-Salcedo et al. and 1\ cm$^{-3}$ for Cloudy. The 
diffuse  intercloud gas thus moves from the cold phase to the warm phase. The 
decrease in the cold gas mass fraction is most significant for the Rosen \& 
Bregman cooling function. Only the ``GMC'' gas which is 70\% of the total, 
remains in the cold phase. For the Sanchez-Salcedo et al., resp. Cloudy, cooling function, 
some of the gas surrounding the clouds is also in the cold phase so that the 
mass fractions are 85\%, resp. 96\%.   The mass fraction of gas that is in the 
cold phase (i.e. molecular gas and CNM) in the inner Galaxy disk region of the molecular ring 
is $\sim 0.5$ (Wolfire et al. 2003). The simulation values for this mass fraction are 
somewhat higher, which we expect is due mostly to their present lack of ionization and 
supernova feedback processes.

\begin{figure}
\begin{center}
\includegraphics[width=8cm]{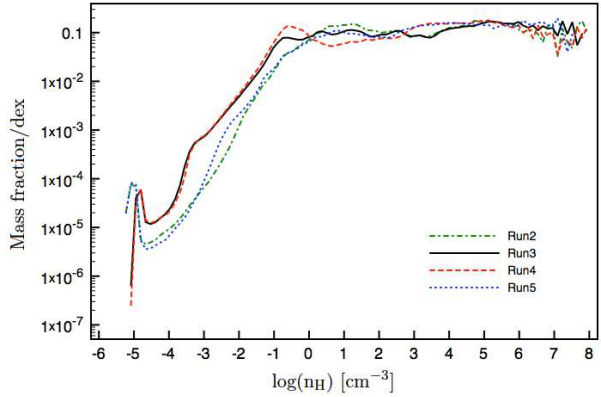}
\caption{Same as Fig.~7, but for Run 2 (green, dash-dotted), Run 3 (solid), 
Run 4 (red, dashed) and Run 5 (blue, dotted).}
\label{fig:PDF_multiphase}
\end{center}
\end{figure}

Together with the temperature, the pressure in the intercloud region increases significantly. 
For example, for $n_{\rm H} = 1$\ cm$^{-3}$, the pressure difference is of the order 
10$^4 k_B$, where $k_B$ is the Boltzmann constant.  The associated heating time scales 
are of the order of 0.1 Myr for 0.1\ cm$^{-3}$, but are much shorter and even below the 
numerical time step for $n_{\rm H} > 1$\ cm$^{-3}$.  (Remember that the 
time step is determined by the Courant condition for the hydrodynamics and not limited 
by the cooling time.) As a consequence, a significant amount of energy is added 
during the first time step to the simulation, i.e. of the order of 10$^{50}$\ erg for 
the Rosen \& Bregman and the Sanchez-Salcedo et al. functions and 10$^{48}$\ erg for the 
Cloudy function.  Note that this initial adjustment is unphysical, and can 
be regarded as a transient associated with the initial conditions. The added energy 
is primarily deposited near the midplane of the disk and eventually causes the disk to 
expand. The mean disk thickness for Run 3 and Run 4 (the atomic cooling functions) is 
$\sim$ 600~pc, an increase of a factor of 4, after 10\ Myr. Because of the lower energy 
deposit for the Cloudy function, the mean disk thickness of Run 5 increases only by 30\% 
to 190~pc.  The larger disk of Run 3 and 4 can also be seen in Fig.~\ref{fig:PDF_multiphase} 
where the PDFs show that the densities between 10$^{-4}$ and 1~cm$^{-3}$ contain more mass 
than in the simulations without heating.

While the higher external pressure causes the disk to inflate, it also acts 
as an additional force confining the molecular clouds.  The higher external pressure resulting 
from this heating of the disk is, however, much smaller than the internal pressure of the 
GMCs, which is set by their self-gravitating weight.  

\begin{figure}
\begin{center}
\includegraphics[width=8cm]{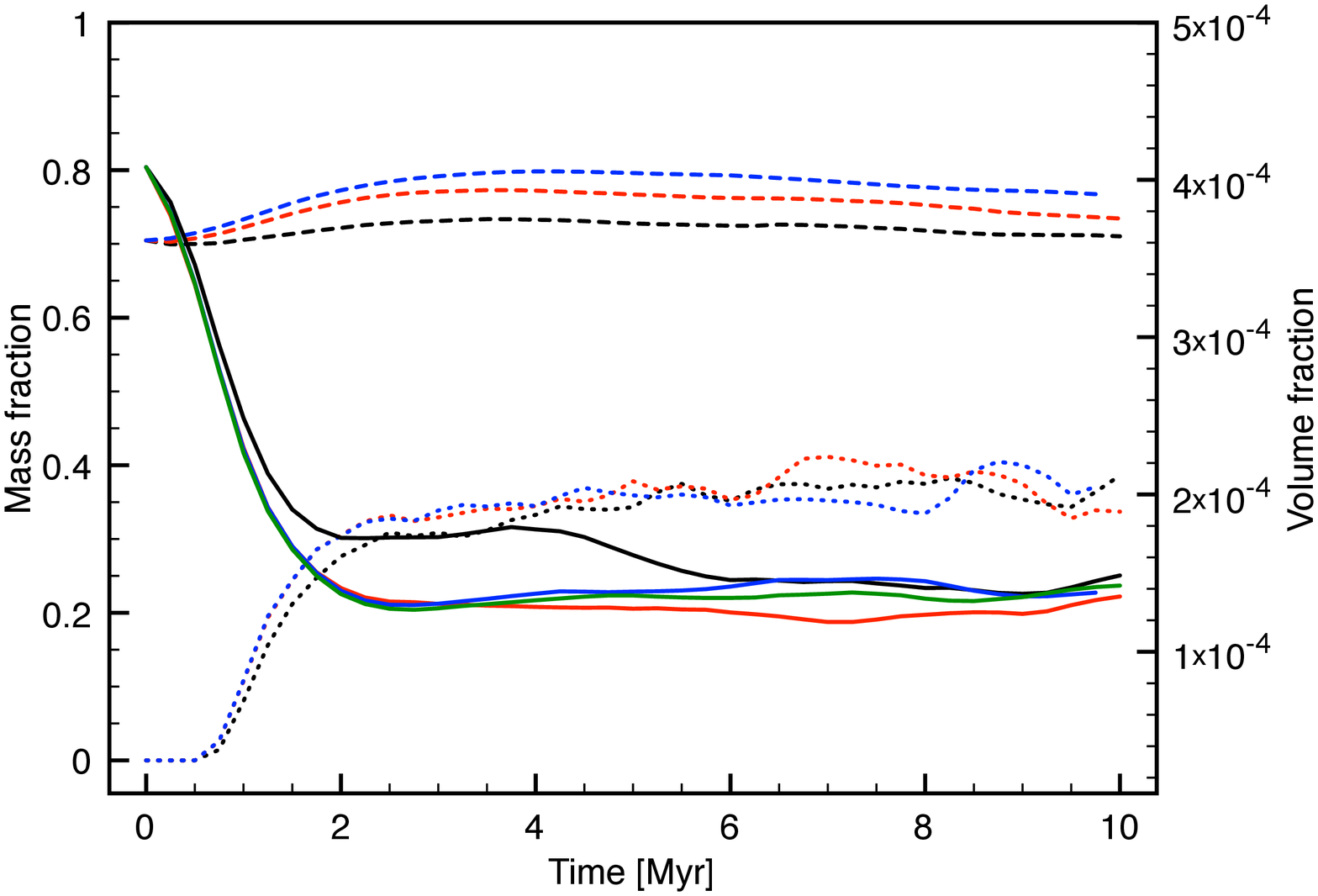}
\caption{Same as Fig.~8, but for Run 3 (black), Run 4 (red) and 
Run 5 (blue).   We also plot the volume fraction for Run 6 (green). The mass 
fraction for Run 6 is plotted in Fig.~\ref{fig:sfr_mfrac}.}
\label{fig:cooling_mfrac} 
\end{center}
\end{figure}

For the runs with the Sanchez-Salcedo et al. and Cloudy cooling functions, the gas in the 
interior of the clouds cools and loses pressure.  This, potentially, has a large effect on 
the cloud evolution. However, by comparing the volume fraction of ``GMC'' gas 
in Run 4 and 5 (when the cloud loses internal thermal pressure)  with the one in Run 3 
(where the internal thermal pressure of the cloud stays constant), we find that the resulting 
effect is minimal (see Fig.~\ref{fig:cooling_mfrac}). The mass distributions above 
10$^2$\ cm$^{-3}$ are nearly identical with a small increase to 75\% for Run 4 and to 79\% 
for Run 5 (see Figs.~\ref{fig:cooling_mfrac} and \ref{fig:PDF_multiphase}). Also, the amount 
of gas that ends up in dense clumps is independent of the cooling function. Furthermore, 
the clouds found in Run 3, 4 and 5 after 10 Myr using the cloud-finding algorithm have 
similar masses, velocity dispersions and sizes. 

Since the gas in the clouds is cold and predominantly molecular, not atomic, 
we also ran a simulation with the Cloudy cooling function and where we use $\mu = 2.33$ 
instead of $\mu = 1.27$. This change does not affect the dynamics of the clouds as can 
be seen in Figs.~\ref{fig:cooling_mfrac}, \ref{fig:PDF_sfr} and \ref{fig:sfr_mfrac}.

\begin{figure}
\begin{center}
\includegraphics[width=8cm]{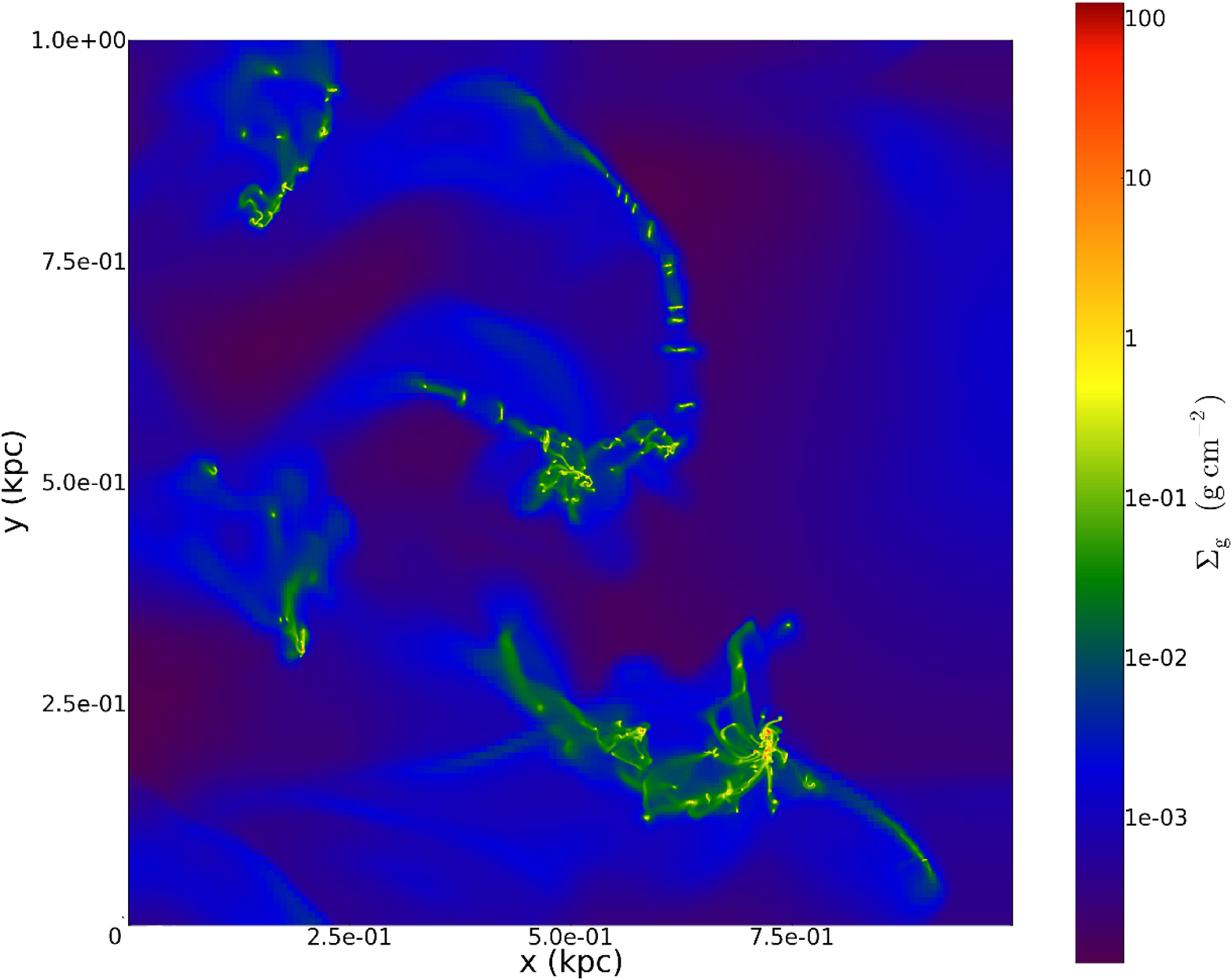}
\includegraphics[width=8cm]{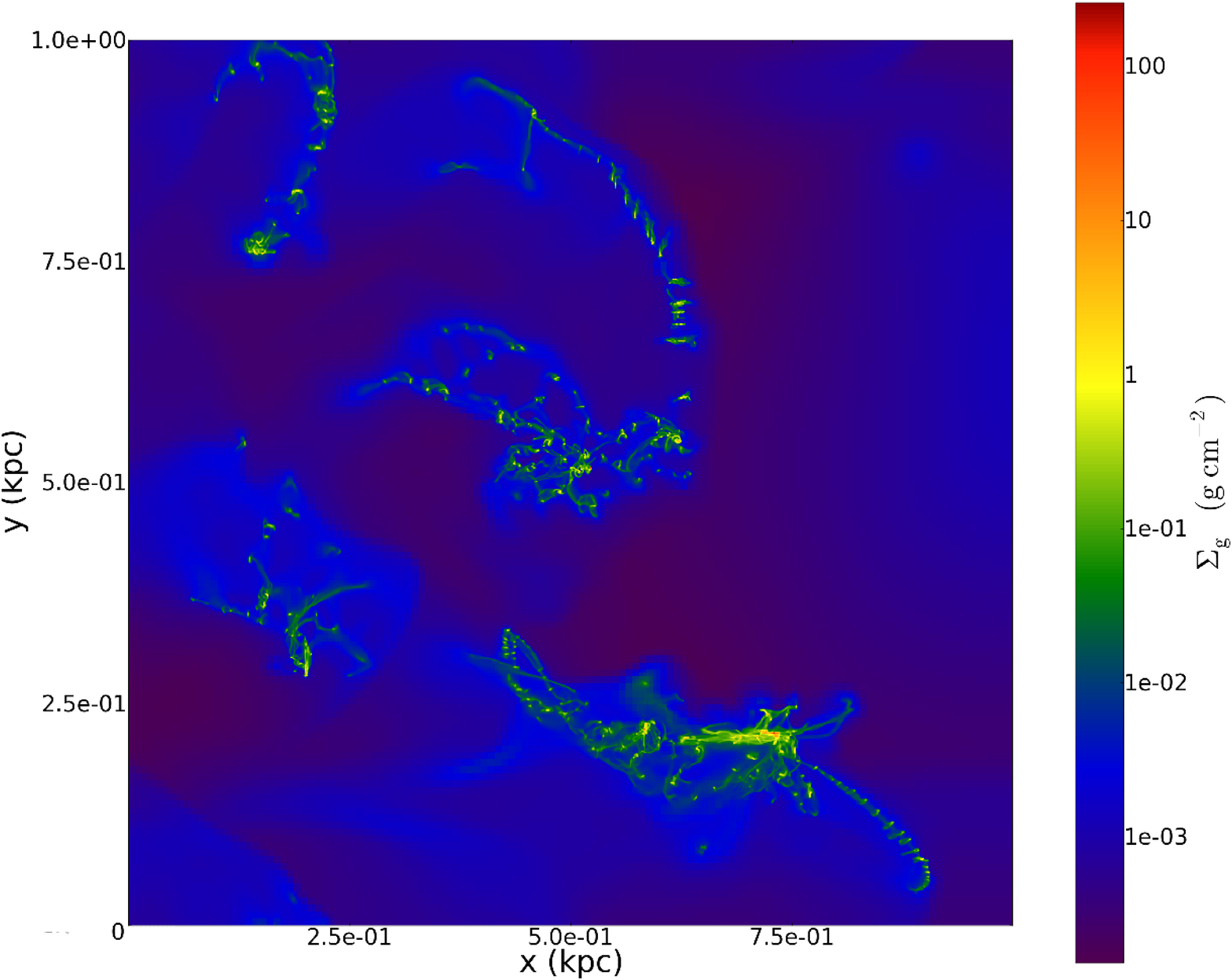}
\includegraphics[width=8cm]{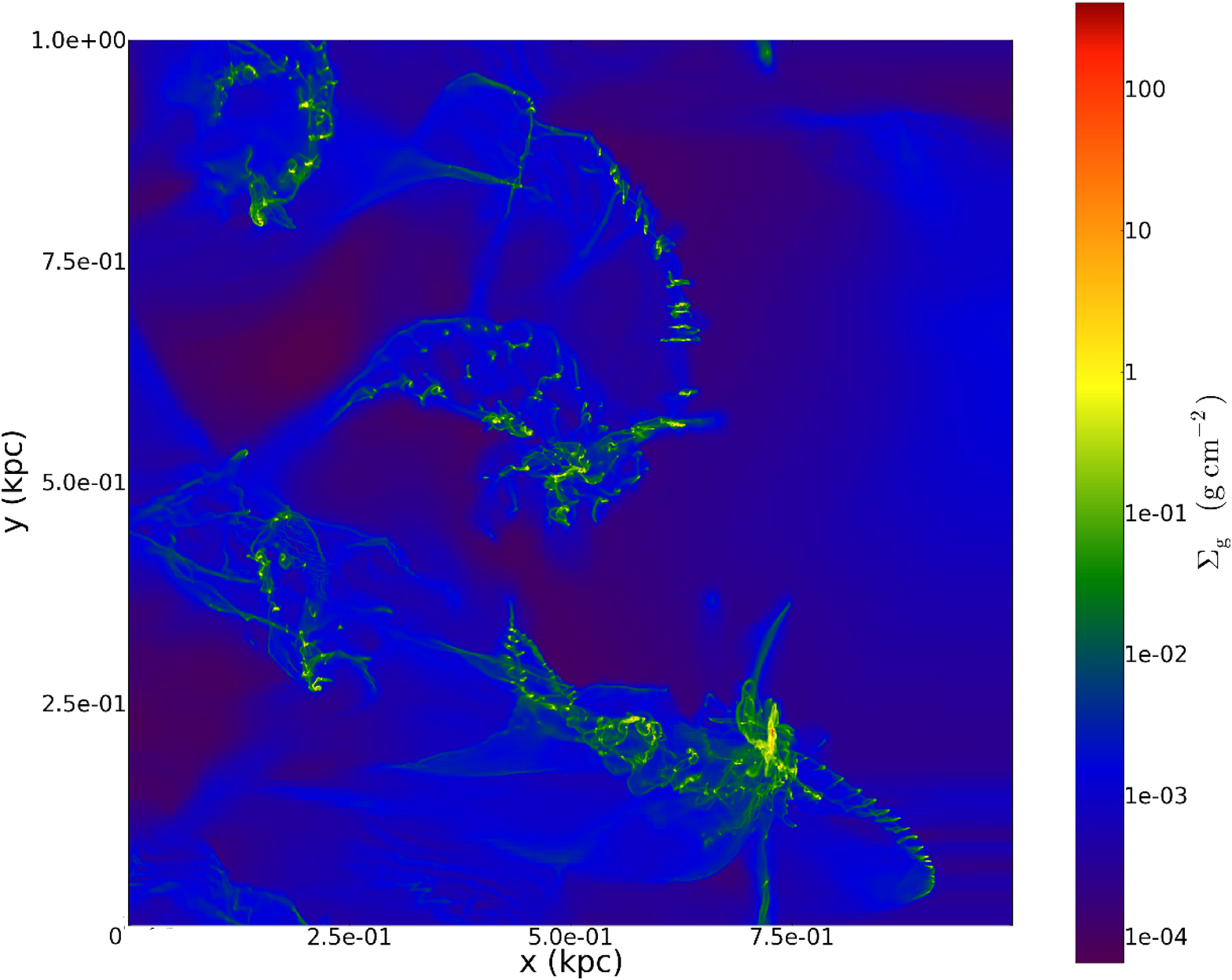}
\caption{Same as Fig.~6 but for Run 3 (top left), Run 4 (top right) and Run 5 
(bottom).}
\label{fig:multiphase}
\end{center}
\end{figure}

While the global properties of the clouds are similar, the cloud substructure, i.e. the 
clump distribution, changes with the cooling function (see Fig.~\ref{fig:multiphase}). Much 
more filamentary and clumpy structures are present for the Cloudy cooling function than 
for the other two. However, this is partly a numerical effect due to 
the applied refinement criterion and to the timestep used for evolving the simulation.  
As the different cooling functions tend to cool the gas to different equilibrium 
temperatures (see Fig.~\ref{fig:RBvsSS}), the densities at which a grid cell is refined 
differ as the gas temperature influences the Jeans length. Such changes introduce small 
variations in AMR simulations \citep[][]{Niklausetal2009}.  

\subsection{Star Formation}
The high resolution simulations described above (Runs 2 - 6) evolve to 
a state where a large fraction, ~40-50\%, of the gas is in clumps, as defined by 
$n_{\rm H} > 10^5~{\rm cm^{-3}}$. Now, in Run 7, we introduce star formation in these 
objects, following the method described in \S 2.2. As the density threshold for gas 
in dense clumps is the same as the critical density of our star formation routine, 
star particles will form in the dense clumps. 

\begin{figure}
\begin{center}
\includegraphics[width=8cm]{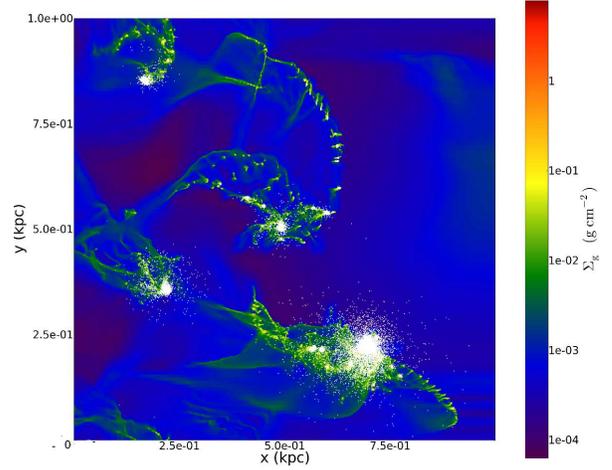}
\caption{Same as Fig.~6 but for Run 7. The white dots represent the star 
particles.}
\label{fig:sfr_cl}
\end{center}
\end{figure}

\begin{figure}
\begin{center}
\includegraphics[width=8cm]{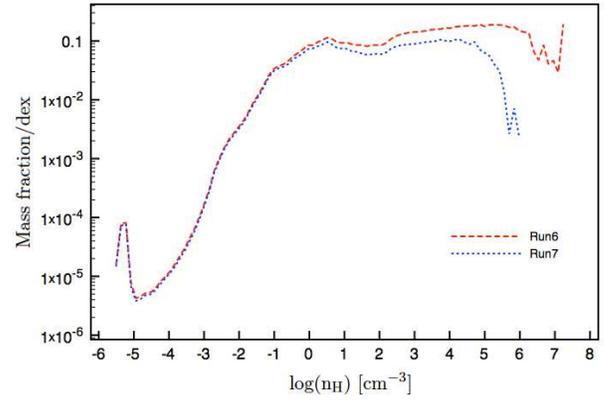}
\caption{Mass-weighted PDF for Run 6 (red, dashed) and Run 7 (blue, dotted). We included the 
stellar mass in the total mass to normalize the distribution function.}
\label{fig:PDF_sfr}
\end{center}
\end{figure}

Figure~\ref{fig:sfr_cl} shows the gas surface density of Run 7 with the star cluster 
particles plotted on top. The star cluster particles are concentrated within the molecular 
clouds. Note, however, that star particles can be ejected from the clouds, especially in 
clouds with large amounts of angular momentum, e.g. from a collision.  The star formation 
has not changed much of the global density structure or dynamics, but has reduced the 
maximum gas surface densities by about an order of magnitude compared to Run 6 without 
star formation.  This can also be seen in the PDFs where the star formation process only
produces a deviation above $n_{\rm H} \approx 10^4$~cm$^{-3}$ and limits the maximum 
density in the simulation to $\approx 10^6$~cm$^{-3}$ (Fig.~\ref{fig:PDF_sfr}). As we do 
not yet include stellar feedback in our simulations, the stars only interact gravitationally 
with their maternal cloud.  

\begin{figure}
\begin{center}
\includegraphics[width=8cm]{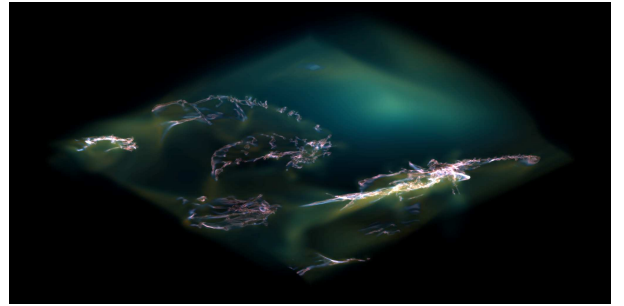}
\caption{Volume rendered number density of Run 7. The ``GMC'' threshold volume 
density, $n_{\rm H}\sim 100\:{\rm cm^{-3}}$, is coloured blue, while the ``Clump'' threshold 
volume density, $n_{\rm H}\sim 10^5\:{\rm cm^{-3}}$, gas is coloured red.}
\label{fig:render}
\end{center}
\end{figure}

\begin{figure}
\begin{center}
\includegraphics[width=8cm]{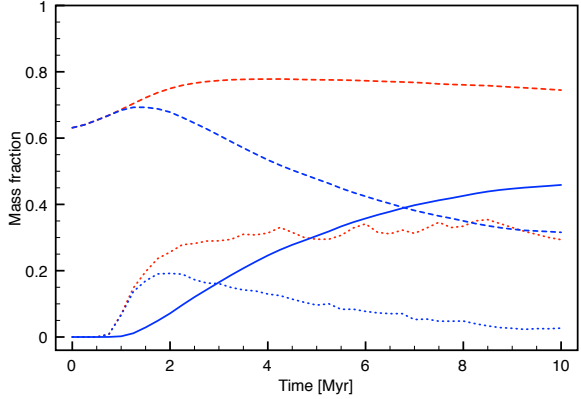}
\caption{Same as Fig.~8, but for Run 6 (red) and Run 7 (blue). 
The stellar mass fraction for Run 7 (solid) is also shown. For Run 7 the 
normalization is done with the sum of the total gas and stellar mass.}
\label{fig:sfr_mfrac} 
\end{center}
\end{figure}

Figure~\ref{fig:render} shows a rendered visualization of the ISM
structures at the end of Run 7, highlighting the volume density thresholds that define 
``GMCs'' and ``Clumps''. Several hundred parsec long filaments of high density gas are 
visible, which bear a qualitative resemblance to some observed Galactic infrared dark cloud
structures \citep[e.g. the ``Nessie'' nebula,][]{Jacksonetal2011}. A comparison of dynamical 
state of the simulation filaments with observed IRDCs \citep[e.g.][]{Hernandezetal2012}, 
will be carried out in a subsequent paper.

\begin{figure}
\begin{center}
\includegraphics[width=8cm]{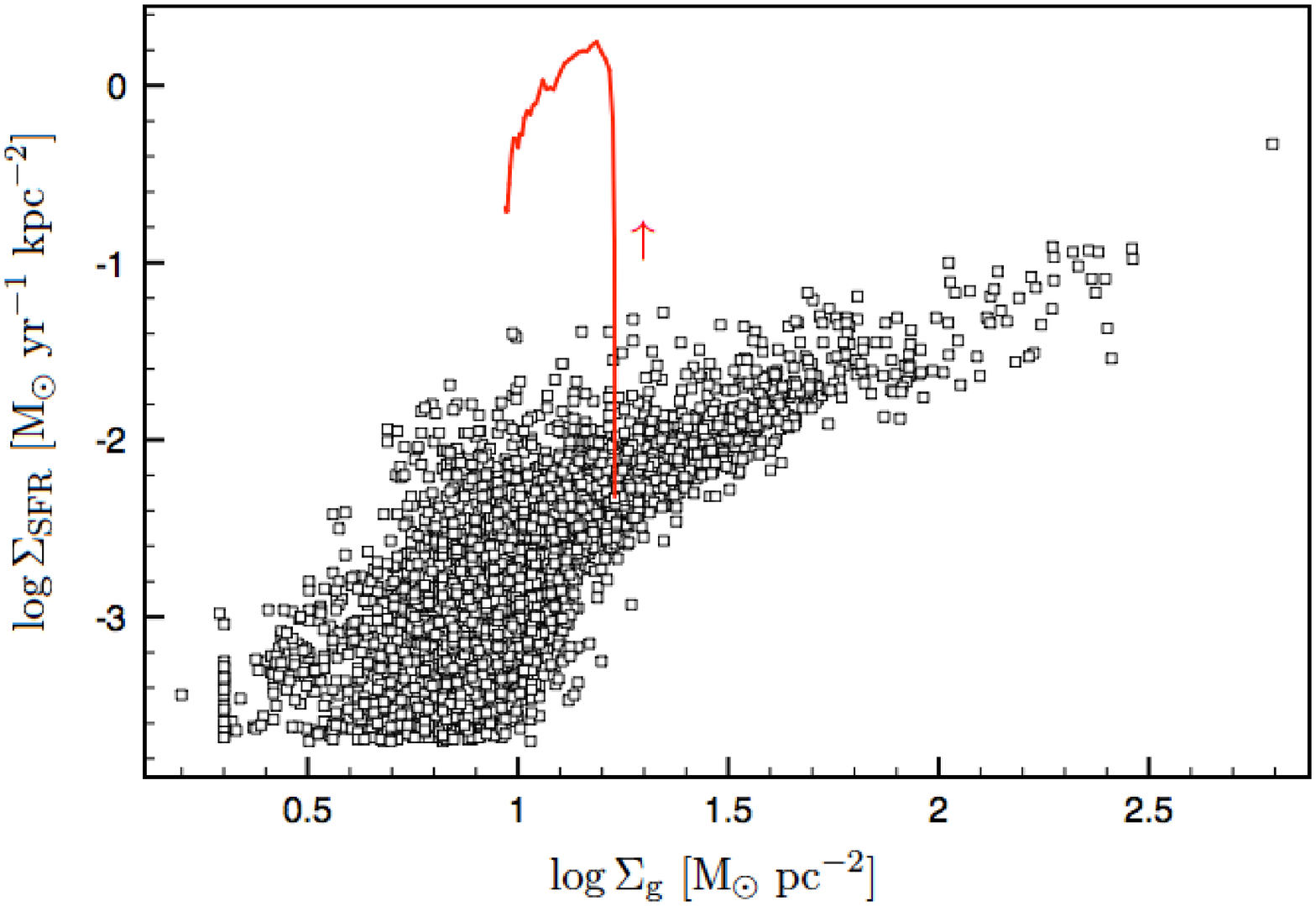}
\includegraphics[width=8cm]{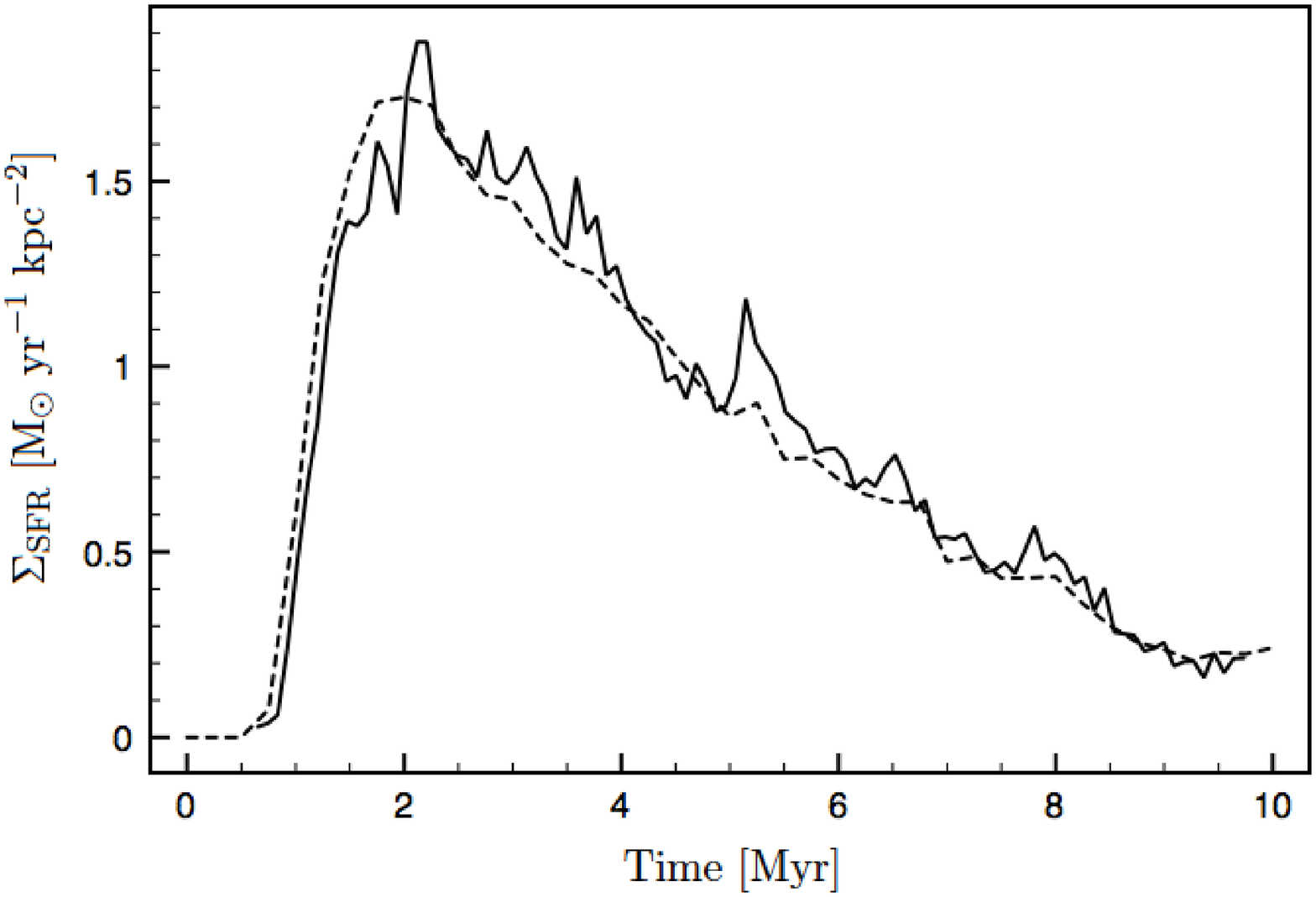}
\caption{{\it Left:} The star formation rate ($\Sigma_{\rm SFR}$) as function 
of the gas surface density. The symbols show the observations of \citet[][]{Bigieletal2008}, 
while the solid line shows the evolution of $\Sigma_{\rm SFR}$ in Run 7. The arrow shows the 
direction of the evolution. The star formation rate quickly rises to more than 100 times the 
observed value --- a result both of the initial condition having a relatively high mass 
fractions of dense gas and the lack of support from magnetic fields or disruption by stellar 
feedback. Gas is consumed, but the SFR remains about 100 times larger than the levels seen in 
galactic disks with similar gas content.  {\it Right:} $\Sigma_{\rm SFR}$ (solid) as 
function of time for Run 7. The dashed line shows the rescaled mass fraction of gas in dense 
clumps for Run 7.}
\label{fig:sfr} 
\end{center}
\end{figure}

The removal of mass from the gas phase because of star formation can be seen by following 
the gas mass fraction in molecular clouds and in dense clumps for Run 6 and 7 
(Fig.~\ref{fig:sfr_mfrac}). After 2\ Myr, dense clumps start to form and immediately produce 
stars. As molecular gas is converted into stars, the mass fraction of gas in molecular clouds 
starts to decrease (Note that we calculate the mass fraction to the total mass including the 
stellar mass). After 10\ Myr, the mass fraction in ``GMCs'' in Run 10 is about 60\% lower 
than in Run 6. Similarly, the dense clump mass fraction drops to 2\% of the total mass compared 
to 32\% in Run 6. As the stellar mass is about half of the total mass after 10\ Myr and star 
cluster particles only originate within dense clumps, this suggests that the dense clump gas 
is continuously replenished from the lower density molecular gas.  This replenishment, however, 
does not keep up with the rate at which the dense clumps convert gas into stars. The free-fall 
time of gas within the clump is shorter than the free-fall time of the region surrounding the 
clump, i.e. the mean density decreases when including the gas around the clump. Hence, the 
mass fraction in dense clumps decreases with time. As the star formation rate depends on the gas
mass in dense clumps (Eq.~\ref{eq:mstar}), the evolution of the star formation rate should 
follow the curve of the mass fraction of the dense clumps. Figure~\ref{fig:sfr} indeed shows 
that this is the case with a maximum star formation rate of 1.8\ M$_\sun$ yr$^{-1}$ kpc$^{-2}$ 
after 2\ Myr and then a steady decrease to 0.2\ M$_\sun$\ yr$^{-1}$  kpc$^{-2}$ at 10 Myr. 
This is more than 2 orders of magnitude more than the star formation rates observed 
by \citet[][]{Bigieletal2008} (see Fig.~\ref{fig:sfr}).

Of course, the star formation rates in Run 7 depend on the values of the parameters 
$\epsilon_{\rm ff}$ and potentially on M$_{\rm *,min}$ and $n_{\rm H,sf}$ in the star 
formation routine. We did additional simulations where we varied these parameters.  Increasing 
M$_{\rm *,min}$ to 200\ M$_\sun$ shortens the run time of the simulation as the number of 
star particles in the simulation decreases, but it does not change the star formation rate. 
The total stellar mass only deviated by less than 2\%. By changing $n_{\rm H,sf}$ to 
10$^4$\ cm$^{-3}$, stars start to form earlier as these densities are reached at earlier times. 
However, the overall star formation rate is not affected very much: the total stellar mass only 
increases by 6\%. On the other hand, if we increase $n_{\rm H,sf}$ to 10$^6$\ cm$^{-3}$ the
stellar mass decreases by $\approx$ 26\%.  The increased critical density reduces the gas 
reservoir from which the star particles form, although only by a relatively small amount.

\section{Conclusion and discussion}\label{sect:discussion}
We have investigated the star formation process down to $\lesssim$parsec scales in a galactic 
disk. We extracted a kiloparsec-scale patch of the disk from the large, global simulation of 
TT09 and increased the resolution down to 0.5~pc. This allowed us to study the structure and
evolution of GMCs in greater detail. We also included additional physics such as heating, 
atomic and molecular cooling, and a simplified approach to star formation. So far we have 
neglected the countering effects of magnetic fields and localized stellar feedback.

We used a novel approach to include molecular cooling in our models. The formation of molecules 
depends strongly on the amount of attenuation of the radiation field. From a high-resolution 
simulation including the atomic cooling function of \citet[][]{SanchezSalcedoetal2002} that 
reproduces the equilibrium phase curve of \citet[][]{Wolfireetal1995} we find that the column
extinction can be expressed as a function of gas density. Such a one-to-one relation eliminates 
the need for time-consuming column-extinction calculations to assess the attenuation of the
radiation field in the numerical simulations. We also use this extinction-density relation 
to generate a table of cooling and heating rates as function of density and temperature with 
the code Cloudy. The resulting cooling function resembles the atomic cooling 
function up to densities of 10$^2$ cm$^{-3}$, above which molecular species start to dominate 
the cooling rates. However, we need to keep in mind that the heating and cooling rates are only 
first order approximations as the extinction law is only a mean relation and as 
local abundance variations and time-dependent chemistry are not considered. Furthermore the 
simulations do not take into account the local generation of FUV radiation from young star 
clusters.

With an increased resolution of $\sim 0.5$ pc our simulations are able to capture a significant 
range of the internal structure of molecular clouds.  While the global properties, such as the 
mass in the molecular clouds, remain the same, filaments and dense clumps form within the 
clouds shifting the mass distribution towards higher densities. The mass distribution within 
the molecular clouds are independent of the applied cooling function even though the three
cooling functions describe different aspects of the thermal properties of the ISM.  This 
suggests that the thermal pressure is of minor importance within the gravitationally-bound clouds. 
Then self-gravity and non-thermal motions determine the cloud structure.  The 
nonthermal motions are driven by bulk cloud motions inherited by the GMCs that were formed 
and evolved in a shearing galactic disk 
where cloud-cloud collisions are frequent and influence GMC dynamics (TT09, 
\citet[][]{Tanetal2012}). Our simulations begin to resolve the cascade 
of the kinetic energy from these larger scales and processes down to the smaller scales of clumps 
in the GMCs.  This approach is to be contrasted with the method of driving turbulence 
artificially in periodic box simulations of GMCs \citep[e.g.][]{Schmidtetal2010} or of 
forming GMCs in large-scale converging flows of atomic gas where the turbulence is driven 
by thermal and dynamical instabilities \citep[e.g.][]{Heitschetal2008,VazquezSemadenietal2011}. 

Of the gas within molecular clouds 50-60\% is in dense clumps with $n_{\rm H}>10^5\:{\rm cm^{-3}}$. 
This value is much higher than observed in nearby GMCs, e.g. 90\% of the clouds in the Bolocam
Galactic Plane Survey have a ratio of clump mass to cloud mass, or clump formation efficiency, 
between 0 and 0.15 \citep[][]{Edenetal2012}. The high clump formation efficiency 
is partly due to our resolution limit. We do not properly capture the formation of individual PSCs. 
so that the turbulent dissipation range is not fully resolved. The clumps then lack turbulent 
support against self-gravity hereby attaining higher densities and accumulating more mass. 

The surface densities of the clumps in Run 7 are in the range of $\sim$0.1 to $\sim$10~g~cm$^{-2}$.
Galactic IRDCs \citep[e.g.][]{ButlerTan2009,ButlerTan2012} and star-forming clumps 
\citep[e.g.][]{Muelleretal2002} are found in the range $\sim$0.1 to a few g~cm$^{-2}$. The most 
extreme mass surface density clumps seen in our simulations are probably prevented from forming 
in reality by localized feedback processes from star formation, especially the momentum input 
from protostellar outflows \citep[][]{NakamuraLi2007}.

The star formation rate in our simulations exceeds that expected from the Kennicutt-Schmidt 
relation \citep[e.g.][]{Bigieletal2011}. These authors find a close relation between the surface 
density of H$_2$ and the star formation rate surface density for 1\ kpc resolution regions within 
31 disk galaxies. With a surface density of roughly 8\ M$_\sun$\ pc$^{-2}$ in our simulations, 
the star formation rate of 0.3\ M$_\sun$\ yr$^{-1}$\ kpc$^{-2}$ after 10~Myr is about 
100 times higher than observed.  This over efficiency of star formation in our simulations is 
a simple reflection of the high mass fraction in dense clumps:  while the GMCs are 
globally stable, there is no support against free-fall collapse in local regions of the GMCs.
We can identify two physical mechanisms of reducing this mass fraction: magnetic fields and stellar 
feedback.  From \citet[][]{ChandraFermi} we can estimate the critical magnetic 
field required for support against self-gravity (neglecting the contribution of thermal and turbulent 
support), i.e. 
\begin{equation}
	\bar{B}_{\rm crit} = 2\pi R \bar{\rho} \sqrt{G}.
\end{equation}
Assuming that the clumps in a GMC condense out of local volumes of radius $\sim$ 10~pc and a GMC 
density of $n_{\rm H} = 100~{\rm cm^{-3}}$, we find that a mean magnetic field of $\sim$ 10~$\mu$G 
is sufficient. This is similar to the observationally inferred value at this density 
\citep[][]{Crutcheretal2010}. We will study the effect of magnetic support in a 
subsequent paper.

\acknowledgements We thank the anonymous referee for his/her comments that improved 
the paper, Paola Caselli and Sam Falle for useful discussions and Elizabeth Tasker for providing 
the initial conditions of our simulations. SvL acknowledges support from the Theory
Postdoctoral Fellowship from UF Department of Astronomy and College of Liberal Arts and Sciences 
and from the SMA Postdoctoral Fellowship of the Smithsonian Astrophysical Observatory. JCT acknowledges 
support from NSF CAREER grant AST-0645412; NASA Astrophysics Theory and Fundamental Physics grant 
ATP09-0094; NASA Astrophysics Data Analysis Program ADAP10-0110. Resources supporting this work were 
provided by the NASA High-End Computing (HEC) Program through the NASA Advanced Supercomputing 
(NAS) Division at Ames Research Center.

\end{document}